\documentclass[final, 3p, times, amsmath, amssymb, 12pt, onecolumn, notitlepage, eqsecnum, longbibliography, superscriptaddress]{elsarticle}
\usepackage{amsmath}
\usepackage{graphicx}
\usepackage{calc}
\usepackage{braket}
\usepackage{slashed}
\usepackage{wasysym}
\usepackage{dsfont}
\usepackage{amsthm,amsmath,amsfonts,amssymb,verbatim,color}
\usepackage{lipsum}
\usepackage{subfigure}
\usepackage{bm}
\usepackage{epsfig,slashed}
\usepackage{hyperref}
\usepackage{flafter}
\usepackage{booktabs}
\usepackage{makecell} % allows to make several lines in one cell
\usepackage{bbm} % to write identity matrices in an expression
\usepackage[export]{adjustbox} %to include figure in an equation
\usepackage{cancel}
\usepackage{epstopdf}
\usepackage{tikz} 
\usepackage{array,color}	
\usepackage{float}
\usepackage{ulem}

\DeclareMathOperator{\sgn}{sgn}

\newcommand{\bk}{{\bf k}}

\newcommand{\bp}{{\bf p}}
\newcommand{\bq}{{\bf q}}

\newcommand{\br}{{\bf r}}

\newcommand{\hV}{{\hat V}}

\newcommand{\hsigma}{{\hat\sigma}}

\def\holOne{\mathds{1}}

\journal{Annals of Physics}

\begin{document}
	\title{BCS-like disorder-driven instabilities and ultraviolet effects in nodal-line semimetals}	
	
	\author{Siyu Zhu, Sergey Syzranov}
	\address{Physics Department, University of California, Santa Cruz, California 95064, USA}

	\begin{abstract}
		We study the effects of quenched disorder on electrons in a 3D nodal-line semimetal. Disorder leads to significant renormalisations of the 
		quasiparticle properties due to ultraviolet processes, i.e. processes of scattering in a large band of momenta, of the width
		exceeding the inverse mean free path.
		As a result, observables such as the density of states and conductivity
		exhibit singular behaviour in a broad range of disorder strengths, excluding a small vicinity of the singular point.
		We find that, for example, the density of quasiparticle states diverges as a function of the disorder strength $g$
		as $\rho(g,E)\propto |g_c(E)-g|^{-2}|E|$ for $g$ smaller than the critical value $g_c(E)$ and crosses over to a constant 
		for $g$ very close to $g_c(E)$, where $E$ is the quasiparticle energy.
		For certain disorder symmetries, a 3D disordered nodal-line semimetal
		can be mapped to a 2D metal with attractive interactions.
		The described disorder-driven instabilities in such a nodal-line semimetal are mapped to Cooper and exciton-condensation instabilities
		in a 2D metal. For other disorder symmetries, the respective instabilities are similar but not exactly dual.
		We discuss experimental conditions favourable for the observation of the described effects.
	\end{abstract}

	%%%%%%%%%%%%%%%%%%%%%%%%%%%%%%%%%%%%%%%%%%%%%%%%%%%%%%%%%%%%%%%%%%%%%%%%%%%%%%%%%%%%%%%%%%%%%%
	
	\maketitle
	
	%%%%%%%%%%%%%%%%%%%%%%%%%%%%%%%%%%%%%%%%%%%%%%%%%%%%%%%%%%%%%%%%%%%%%%%%%%%%%%%%%%%%%%%%%%%%%%
	
	\section{Introduction}
	
\begin{figure}[h]
		\centering
		\includegraphics[width=0.4\linewidth]{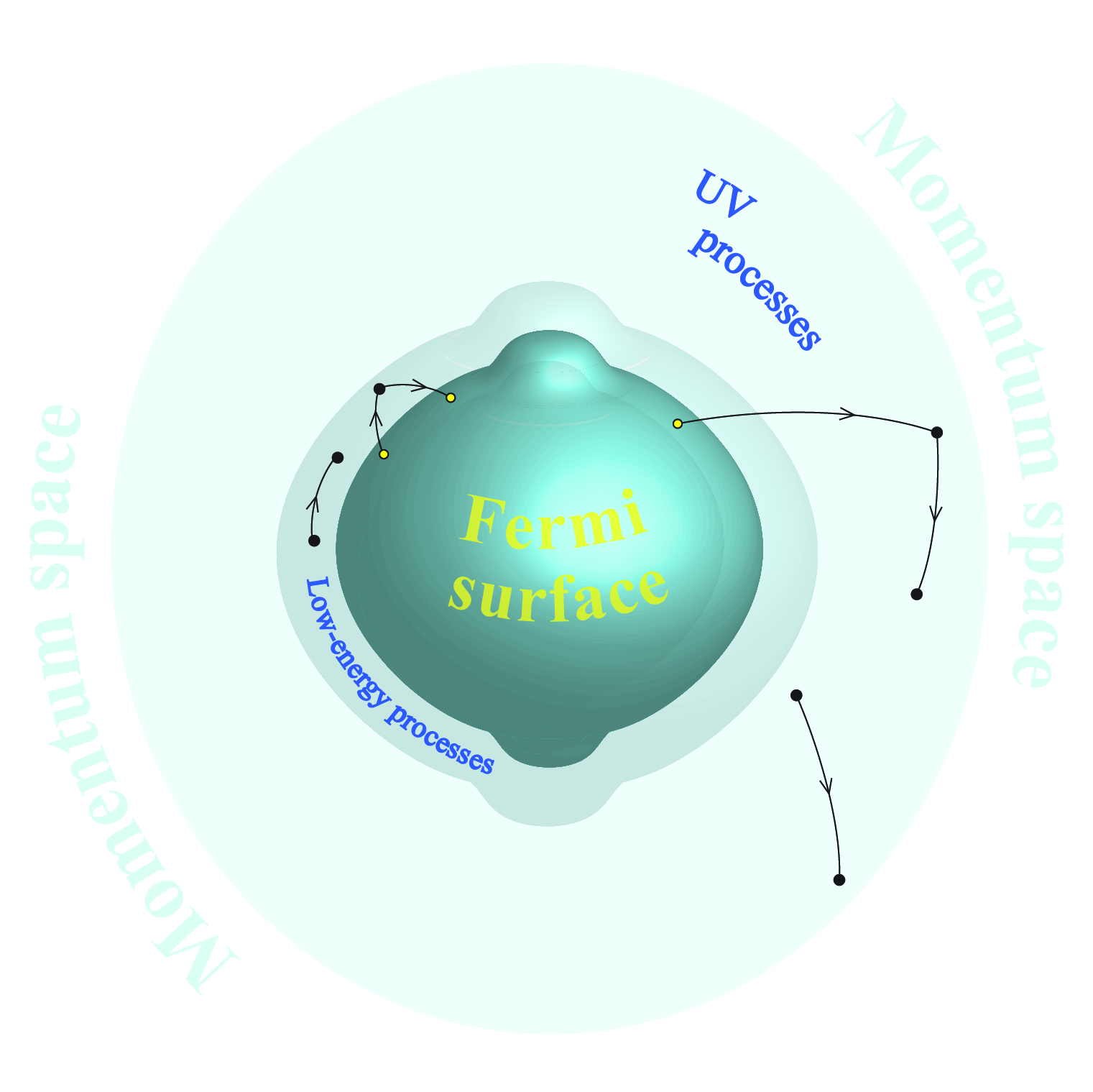}
		\caption{\label{fig:fermisscattering}
			Ultraviolet and low-energy processes of quasiparticle scattering occur, respectively, far and close to the Fermi surface in a conducting material.}
	\end{figure}

	Thermodynamics and transport in conducting materials are usually studied by considering only
	quasiparticle states and collective modes near the Fermi surface.
	Scattering through the states far from the Fermi surface,
	hereinafter referred to as 
	{\it ultraviolet (UV) processes}~\footnote{To our knowledge, the term ``ultraviolet processes'' was first introduced by I.L.~Aleiner and K.B.~Efetov in Ref.~\cite{AleinerEfetov:graphene}},
	is believed to only renormalise quasiparticle parameters (e.g. mass, density of states, and elastic scattering
	time) near the Fermi surface (see Fig.~\ref{fig:fermisscattering}),
	without causing qualitatively new effects. The effective renormalised parameters are measured in experiment.
	The system may then be described, in the spirit of the Fermi-liquid theory,
	by effective low-energy models near the Fermi surface or the surface of a given
	energy $E$, using, e.g., the kinetic equation~\cite{Abrikosov:metals} or non-linear sigma-models~\cite{Efetov:book,Kamenev:book}.
	
	This picture, however, is not applicable if the Fermi surface shrinks, e.g., to a point or a line (as shown in Fig.~\ref{fig:NDLS}) in momentum space.
	Such a scenario can occur in nodal semimetals, where the conduction and valence bands touch in momentum space, as exemplified by Dirac/Weyl and nodal-line semimetals~\cite{Armitage:WeylReview,Burkov:review,Fang:NLSreview}.
	The Fermi surface also shrinks to a point in a single-band system, e.g. a dilute quantum 
	gas or a lightly-doped semiconductor, if the chemical potential lies at the band edge.

	In this case, all (quasi-)particle scattering processes are effectively UV and may give rise to qualitatively new
	effects. For example, it has been demonstrated in Ref.~\cite{AleinerEfetov:graphene} that UV processes in graphene with point scatterers cause large logarithmic corrections 
	$\propto \log(E_0/|\mu|)$ to the conductivity (see also Ref.~\cite{OstrovskyGornyiMirlin:disorderedGrapehen}), distinct from the conventional
	weak-localisation corrections~\cite{GorkovLarkinKhmelnitski:WL}, where $E_0$ is the bandwidth (the UV cutoff for energies).
	In the limit of small $\mu$, such corrections come from all length and energy scales available for the quasiparticles in the band.
	
	UV disorder-driven processes have also been predicted to lead, for example, to non-monotonic dependences of conductivity on disorder strength~\cite{Syzranov:WeylTransition} in 
	3D Weyl semimetals and unconventional Lifshitz tails~\cite{Suslov:rare,Wegner:DoS,Syzranov:unconv}\cite{Syzranov:review,TikhonovIoselevichFeigelman:tails} 
	in high-dimensional semiconductors~\cite{Syzranov:review} and systems with power-law hopping~\cite{RodriguezMalyshev:veryveryFirst,Rodriguez:firstPowerLaw,Malyshev:firstPowerLaw,Garttner:longrange}.
	Furthermore, 
	UV effects can lead to novel unconventional phase transitions.
	3D Weyl/Dirac semimetals, high-dimensional semiconductors and certain systems of ultracold particles~\cite{Monroe:longrange,Islam:longrange,Blatt:chain1,Blatt:chain2} have been predicted~\cite{Fradkin1,Fradkin2,Syzranov:review},
	under some approximations, to display disorder-driven transitions at $\mu=0$ distinct from the conventional Anderson metal-insulator transitions.
	Such UV transitions would lie in different universality classes
	and display, e.g., singular behaviour of the density of states at the transition, which never happens for the conventional 
	Anderson transitions (at $\mu\neq 0$). 
	Exponentially rare UV non-perturbative effects (rare-region effects) may broaden UV disorder-driven criticalities and convert them to sharp crossovers, as discussed recently in the context of 3D Weyl semimetals~\cite{Wegner:DoS,Suslov:rare,Nandkishore:rare}\cite{Syzranov:unconv,PixleyHuse:missedPoint,WilsonPixley:rareSingleCone,PixleyWilson:reviewRare}.
	However, such crossovers may nevertheless appear as phase transitions in experiment, and, in some systems,
	allow for a parametric separation of the UV criticality and rare-region effects~\cite{SbiesrkiSyzranov:scaling}.
	Interacting nodal semimetals also display phase unconventional phase transitions, e.g., superconductive instabilities~\cite{Kozii:superconductivityDirac,MaciejkoNandkishore:WeylWrongRG} that are distinct from those 
	in systems with large Fermi surfaces.
	
	Recently, it has been demonstrated in Ref.~\cite{SunSyzranov:duality} that a broad class of interacting disorder-free systems allows for a mapping to
	disordered non-interacting systems to all orders of the perturbation theory.
	This allows one to predict novel interaction-driven (disorder-driven) phase transitions dual to the previously known
	disorder-driven (interaction-driven) phase transitions and further expands the class of known UV phase transitions.
	In the case of UV disorder-driven transitions converted to crossovers by non-perturbative UV rare-region effects, the dual
	interaction-driven phase transitions may be genuine phase transitions~\cite{SunSyzranov:duality}.
	
	UV disorder-driven phenomena have been
	studied most extensively in the context of 2D (see, for example, Refs.~\cite{LudwigFisher:DiracQHE,AleinerEfetov:graphene,OstrovskyGornyiMirlin:disorderedGrapehen,PanOhtsukiShindou:2DDiracRenormalisation})
	and 3D~\cite{Fradkin1,Fradkin2,GoswamiChakravarty,KobayashiHerbut,SbierskiBrouwer:firstWSM,Syzranov:review,BalogFedorenko:porousMedium}
	Dirac and Weyl materials, systems with long-range power-law hopping~\cite{RodriguezMalyshev:veryveryFirst,Rodriguez:firstPowerLaw,Malyshev:firstPowerLaw,Garttner:longrange} and high-dimensional semiconductors~\cite{Suslov:rare,Syzranov:review}. 
	Another, less explored class of semimetals exhibiting UV disorder-driven effects is nodal-line semimetals, 
	in which the density of states vanishes along a line in momentum space and not at a point as in nodal-point semimetals~\cite{Armitage:WeylReview,Fang:NLSreview}. 
	Such band structures and their manifestations in transport and thermodynamics have been observed, for example,
	in $ZrSiS$~\cite{Schoop:ZrSiS}, $ZrSiSe$~\cite{Shao:ZrSiSe}, $PbTaSe_2$~\cite{BianHasan:PbTaSe},
	$TlTaSe_2$~\cite{BianHasan:TlTaSe}, $NbAs_2$~\cite{ShaoBasov:NbAs2} and $TbSbTe$~\cite{Gao:TbSbTe} and have been theoretically predicted in numerous other materials. 
	Nodal-line semimetals have been numerically predicted, in Ref.~ \cite{Goncalves:CriticalityNLS}, to display non-Anderson disorder-driven criticality between weak-disorder and strong-disorder phases,
	despite the UV renormalisations leading to
	marginal relevance of weak disorder~\cite{WangNandkishore:disorderEffectNLS} similar to that for 2D Dirac particles~\cite{AleinerEfetov:graphene, OstrovskyGornyiMirlin:disorderedGrapehen,Fradkin1}, for which
	critical behaviour is not expected.
	
	In this paper, we study disorder-driven UV effects in nodal-line semimetals.
	We develop an approach that allows us to describe UV effects exactly in the limit of 
	a large ratio of the size of the nodal line to the UV momentum cutoff.
	We demonstrate that the UV disorder-driven effects lead to a
	sharp crossover in the density of states and other observables in a nodal-line semimetal.
	In a large interval of disorder strengths and energies, the density of states, for example,
	displays a singular behaviour given by
	\begin{align}
		\rho(E,g_0)\propto \left[g_c(E)-g_0\right]^{-2}|E|,
		\label{DoSSummary}
	\end{align}
	where $E$ is the quasiparticle energy measured from the nodal line; $g_0$ is the bare disorder strength
	and $g_c(E)$ is a known function of energy, which we obtain in this paper.
	In a small vicinity of the critical point [$g_c(E)\approx g_0$], the density of states as a function 
	of energy crosses over to a constant. 
	
	The described disorder-driven singularities in 3D nodal-line semimetals are similar to the Cooper~\cite{Mahan:book,AGD}
	and exciton-condensation~\cite{KeldyshKopaev:ExcitonicCondensation,LozovikYudson:ExcitonicCondensation}
	instabilities in interacting disorder-free 2D metals.
	Furthermore, for certain disorder symmetries, 
	there exists an exact duality mapping~\cite{SunSyzranov:duality} between the model of a disordered nodal semimetal considered in this paper and the models of interacting metals exhibiting Cooper and excitonic instabilities. 
	
	The paper is organised as follows.
	We describe the model of a nodal-line semimetal in the presence of quenched disorder in Sec.~\ref{Sec:Model}.
	The effect of disorder on the quasiparticle properties in such a semimetal is described in 
	Sec.~\ref{Sec:ParticleRenormalisationDisorder}. The behaviour of the quasiparticle density of states is described
	in Sec.~\ref{Sec:DoS}.
	In Sec.~\ref{sec:duality}, we discuss a duality mapping between 3D disordered nodal-line semimetals and interaction-driven phase transitions
	in 2D metals.
	We conclude in Sec.~\ref{Sec:conclusion}.
	
	%%%%%%%%%%%%%%%%%%%%%%%%%%%%%%%%%%%%%%%%%%%%%%%%%%%%%%%%%%%%%%%%%%%%%%%%%%%%%%%%%%%%%%%%%%%%%%%%%%%5
	
	\section{Model}
	
	\label{Sec:Model}
	
	\begin{figure}[H]
		\begin{center}
			\includegraphics[width=0.7\linewidth]{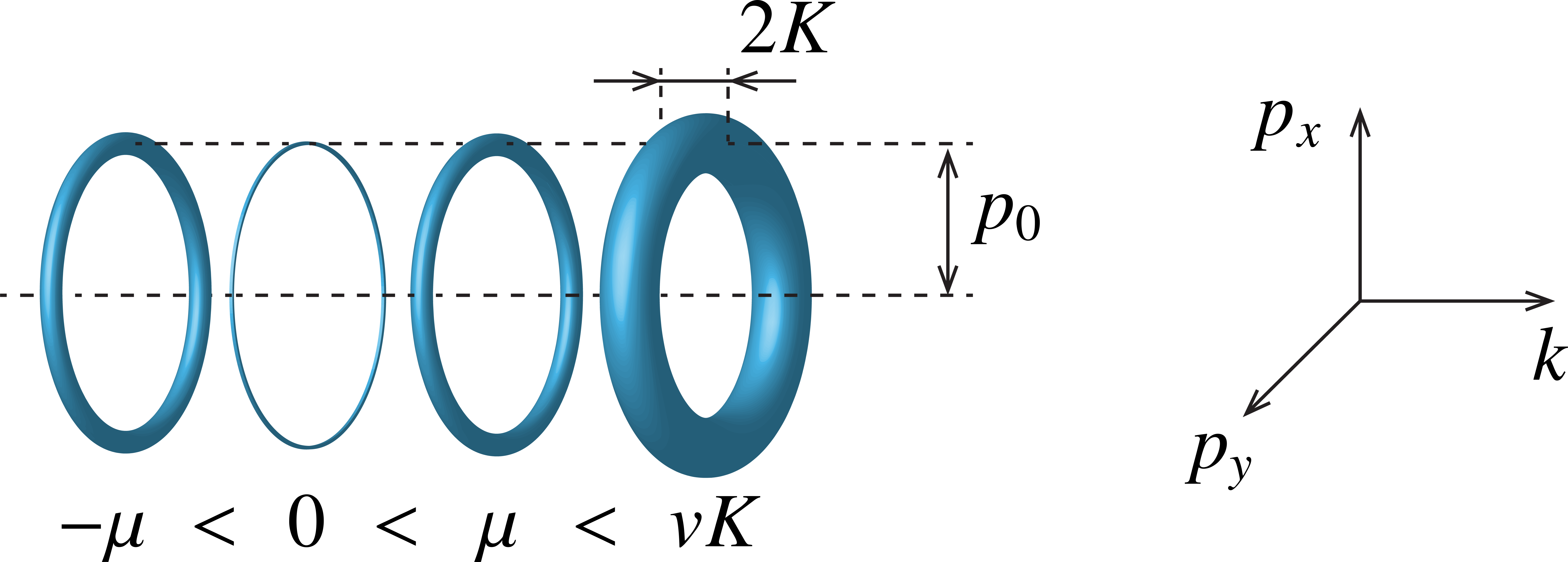}
			\caption{
				\label{fig:NDLS}
				The Fermi surface in a disorder-free nodal-line semimetal for various values of the chemical potential. At zero chemical potential, the Fermi surface shrinks to a line.		
			}
		\end{center}
	\end{figure}
	We consider a nodal-line semimetal with a circular nodal line, shown in Fig.~\ref{fig:NDLS}.
	The Hamiltonian of quasiparticles near the nodal line is given by 
	\begin{align}
		\label{H}
		\hat{H} =  v \left(|\bp|-p_0\right)\hat{\sigma}_{x}
		+v k\hat{\sigma}_{z}
		+\sum_i \hV(\br-\br_i),
	\end{align}
	where $\bp=(p_x,p_y)$ is the 2D vector of the quasiparticle momentum in the $xy$ plane, in which the nodal line ($|\bp|=p_0$) lies (see Fig.~\ref{fig:NDLS}); $k$ is the momentum component along the $z$ axis;
	$\hsigma_{x}$ and $\hsigma_{z}$ are Pauli matrices corresponding to a spin-$1/2$ degree of freedom (pseudospin) in the subspace of the two bands 
	in the nodal-line semimetal; 
	$v$ is the quasiparticle velocity in the directions perpendicular to the nodal line.
	The last term in Eq.~\eqref{H} accounts for impurities with random locations $\br_i$.
	The perturbation $\hV$ created by one impurity may have a structure in the pseudospin space.
	The main focus of this paper is the case of potential disorder, i.e. impurities that do not
	couple to the pseudospin degree of freedom.
	
	{\it Ultraviolet cutoff.} 
	We assume that all quasiparticles are confined to a tube of radius $K\ll p_0$ around the nodal line in momentum space (see Fig.~\ref{fig:NDLS}).
	The processes of scattering beyond that tube may be suppressed
	due to sufficiently fast (faster than linear) growth of the quasiparticle
	energies away from the tube. Also, processes of scattering through states far from the nodal line may be assumed to only renormalise the quasiparticle parameters within the tube without causing qualitatively new effects.

	{\it The ultraviolet momenta} are defined in what follows as the momenta
	$\bp$ and $k$
	that are confined to the nodal tube
	of radius $K$, on the one hand, and that are sufficiently far from the Fermi momentum $k_F=|\mu|/v$,
	further than the inverse mean free path $\ell$, on the other hand:
	\begin{align}
		\ell^{-1}\ll |k_F-k|, \Big\lvert k_F-\big\lvert p_0-\lvert \bp\rvert \big\rvert \Big\rvert<K.
	\end{align}

	%%%%%%%%%%%%%%%%%%%%%%%%%%%%%%%%%%%%%%%%%%%%%%%%%%%%%%%%%%%%%%%%%%%%%%%%%%%%%%%%%%%%%%%%%%%%%%%%%%%%%%%%%%%%%%%%%%%

	\section{Renormalisation of quasiparticle properties by disorder}

	\label{Sec:ParticleRenormalisationDisorder}
	
	\subsection{Types of scattering processes}
	
\begin{figure}[h]
		\centering
		\includegraphics[width=0.7\linewidth]{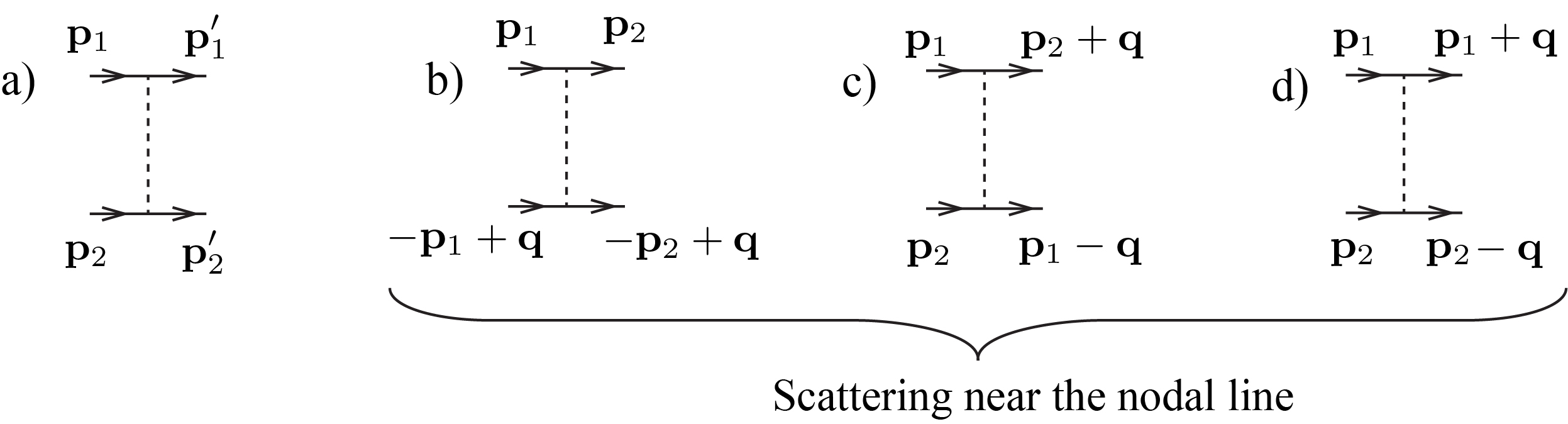}
		\caption{\label{fig:DisorderChannels} Pair-wise correlators 
			of the random potential (``impurity lines'') for a disordered nodal-line semimetal. The transverse quasiparticle 
			momenta (momenta in the plane of the nodal line) are shown.
			(a) A generic impurity line.
			(b)-(d) Possible scattering of quasiparticles near the nodal line, where the momenta $\bp_1$ and $\bp_2$ are close to the nodal line ($|\bp_1|\approx|\bp_2|\approx k_0$),
			and the momentum $\bq$ is significantly smaller than the radius of the nodal line $(|\bq|\ll p_0)$.}
	\end{figure}

\begin{figure}[h]
		\centering
		\includegraphics[width=0.6\linewidth]{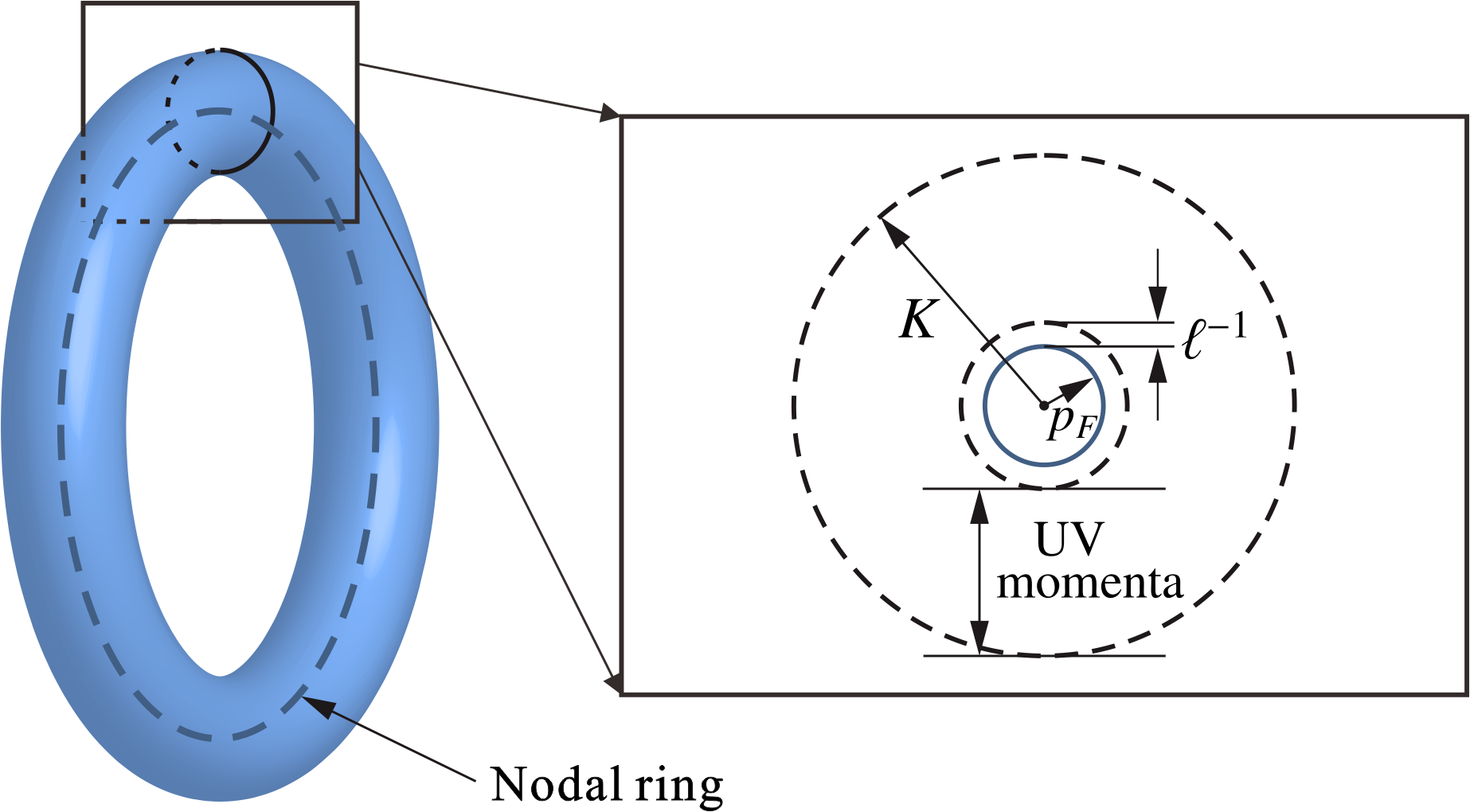}
		\caption{The hierarchy of momenta near the nodal line. $p_F$ is the Fermi momentum measured from 
			the nodal line. The UV processes come from the momenta between $p_F$ and $K$ and lead to a significant
			renormalisation of the quasiparticle properties in the layer of momenta of the width $\ell^{-1}$
			around the Fermi surface. Quasiparticles in this layer of momenta can be described by low-energy theories
			(kinetic equation, nonlinear sigma-models, Fermi-liquid theory, etc.), which use the UV-renormalised quasiparticle
			parameters as input.
			\label{fig:nodalringscattering}
		}
	\end{figure}

	The disorder potential is assumed, for simplicity, to have Gaussian statistics.
	The properties of disorder-averaged quantities are then determined by the ``impurity line'', 
	pair-wise correlators of the random part of the Hamiltonian~\eqref{H}, which are mimicked by the diagram in Figs.~\ref{fig:DisorderChannels}a.
	Both the incoming and outgoing momenta in the impurity line have to be close to the nodal line in momentum space, taking into account the smallness of the UV
	cutoff $K \ll p_0$ measured from the nodal line (see Fig.~\ref{fig:nodalringscattering}).
	
	Momentum conservation law $\bp_1+\bp_2=\bp_1^\prime+\bp_2^\prime$ for the transverse quasiparticle momenta (i.e. momenta in the plane of the nodal line) and the confinement of those momenta to the vicinity of the nodal line restrict possible types of scattering processes.
	As a result, there are three possible types of impurity lines corresponding to: 1) 
	$\bp_1\approx -\bp_2$, $\bp_1^\prime\approx -\bp_2^\prime$
	%$|\bp_1+\bp_2|, |\bp_1^\prime+\bp_2|\lesssim K\ll p_0$
	(``Cooper channel''), 2) $\bp_1\approx \bp_2^\prime$, $\bp_2\approx\bp_1^\prime$ (``excitonic channel'')
	and 3) $\bp_1\approx \bp_1^\prime$, $\bp_2\approx\bp_2^\prime$,
	where $\bp_1$, $\bp_2$, $\bp_1^\prime$ and $\bp_2^\prime$ are the transverse momenta of the quasiparticles, i.e. momenta in the plane of the nodal line (the $xy$
	plane in Fig.~\ref{fig:NDLS}).
	The respective impurity lines imply no constraints on the longitudinal momenta (i.e. momenta along the $z$ axis).
	These impurity lines 1)-3) are shown, respectively, in Figs.~\ref{fig:DisorderChannels}b-d. 
	
	The confinement of the quasiparticle momenta to the vicinity of the nodal line leads to the suppression
	of diagrams with crossed impurity lines by the small parameter
	\begin{align}
		K/p_0 \ll 1,
		\label{SmallParameterNodalLine}
	\end{align}
	for quasiparticles with sufficiently weak elastic scattering rates, $\ell^{-1}\ll K$. %\ttr{how this connect to the scattering rate and how we get $\ell^{-1}\ll K$ here}.
	This suppression is similar to the suppression of the diagrams with crossed impurity lines in conventional metals by the 
	large Ioffe-Regel parameter $p_F \ell \gg 1$~\cite{AGD}.
	Such diagrams in conventional metals are dominated
	by the corresponding vicinity of the Fermi surface ($|p-p_F|\lesssim 1/\ell$) or are not qualitatively affected by the UV processes, which are thus neglected.
	As we show below, the contributions of the UV momenta 
	($\ell^{-1}\ll|p-p_0|\ll K$) in nodal-line semimetals
	can lead to sharp dependencies of observables (density of states, conductivity, specific heat, etc.) on 
	the disorder strength and chemical potential, which may appear as phase transitions in experiment.

	For quasiparticles with large scattering rates ($\ell^{-1}\gtrsim K$), the band of the UV momenta (shown in Fig.~\ref{fig:nodalringscattering})
	does not exist, and the associated UV effects are absent. 
	For such quasiparticles, diagrams with crossed impurity lines are suppressed by large effective Ioffe-Regel parameter $p_0 \ell\gg1$, similarly to the case of a conventional metal~\cite{AGD}.

	%%%%%%%%%%%%%%%%%%%%%%%%%%%%%%%%%%%%%%%%%%%%%%%%%%%%%%%%%%%%%%%%%%%%%%%
	
	\subsection{Ultraviolet renormalisation of disorder}
	
	\label{sec:UVrenormalisationMainResults}
	
\begin{figure}
		\centering
		\includegraphics[width=\linewidth]{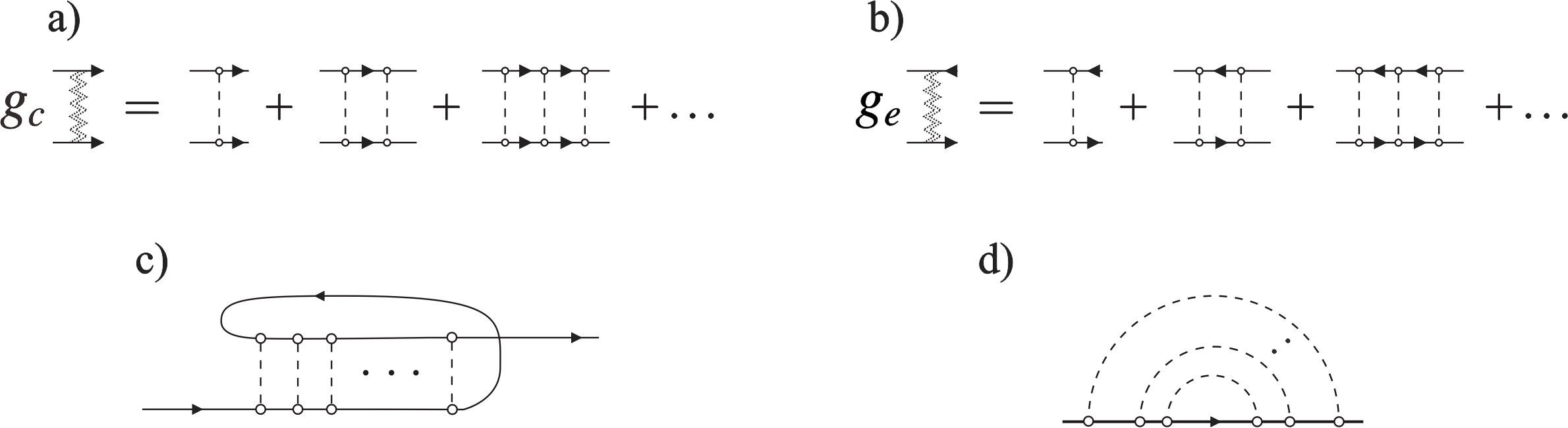}
		\caption{ 
			\label{fig:quasiparticlerenormalisation}
			Diagrams for the renormalisation of the properties of low-energy quasiparticles. a) The ``Cooper channel'' of disorder scattering,
			which corresponds to the process in Fig.~\ref{fig:DisorderChannels}b. %\ttr{here should be Fig3b}
			b) The ``exciton instability channel'' corresponding to the process
			in Fig.~\ref{fig:DisorderChannels}c.
			c-d) The dominant contributions to the self-energy from the respective scattering channels.
		}
	\end{figure}

	The scattering of low-energy quasiparticles, i.e. quasiparticles close to the nodal line, is renormalised
	by the processes of scattering through momentum states far from the nodal line.
	The dominant processes that contribute to the renormalisation of the impurity lines in Figs.~\ref{fig:DisorderChannels}b and \ref{fig:DisorderChannels}c are shown, respectively, 
	in Figs.~\ref{fig:quasiparticlerenormalisation}a and \ref{fig:quasiparticlerenormalisation}b.
	The other processes are suppressed by the small parameter~\eqref{SmallParameterNodalLine}.

	The impurity line renormalised by the processes in Figs.~\ref{fig:DisorderChannels}b (Cooper channel) is given by 
	\begin{align}
		{g}_{c} = g_0 \holOne \otimes \holOne 
		+ g_0^2 \int \frac{d\bp}{(2\pi)^2}\frac{d k}{(2\pi)}G_0\left(k,\bf{p}\right)\otimes G_0\left(-k,-\bf{p}\right)+\ldots,
		\label{GenericExpressionImpurityLine}
	\end{align}
	where 
	$g_0$ is the bare value of the impurity line;
	$\otimes$ stands for the product of the two subspaces corresponding to the two ends of the impurity line, and 
	$G_0(k,\bp) = \left[E-v \left(|\bp|-p_0\right)\hat{\sigma}_{x}
	-v k\hat{\sigma}_{z}\right]^{-1}$.
	The integration in Eq.~\eqref{GenericExpressionImpurityLine}
	is carried out over all momenta between the UV cutoff $K$ and a characteristic ultraviolet scale that is determined 
	by the quasiparticle energy $E$ or the quasiparticle scattering rate and that will be discussed below.
	The momentum $k$ belongs to the ultraviolet range (cf. Fig.~\ref{fig:nodalringscattering}) and significantly exceeds the 
	characteristic momenta of the quasiparticles near the Fermi surface whose properties are being renormalised.
	The disorder strength $g_e$ in the ``excitonic channel'' (corresponding to the scattering process in 
	Figs.~\ref{fig:DisorderChannels}c) is renormalised similarly.
	
	Because the UV renormalisation processes we consider come from momenta sufficiently far from the Fermi surface,
	the analytic properties of the Green's functions $G_0$ near the poles in Eq.~\eqref{GenericExpressionImpurityLine}
	are not important. In most materials, such UV processes would only renormalise the properties of the quasiparticles
	near the Fermi surface, without leading to qualitatively new phenomena.
	
	We show, however, that the UV renormalisations in a nodal-line semimetal considered here 
	result in a singular behaviour of the 
	disorder strength near a critical value (with the singularity broadened in a small vicinity of that value). This leads to a strong dependence of physical observables, such as conductivity and the density of states, as we presented in this paper. 
	The singularities corresponding to Figs.~\ref{fig:quasiparticlerenormalisation}a and \ref{fig:quasiparticlerenormalisation}b
	are similar, respectively, to the Cooper instability~\cite{Mahan:book,AGD} and the  exciton-condensation instability~\cite{KeldyshKopaev:ExcitonicCondensation,LozovikYudson:ExcitonicCondensation}
	in interacting systems. 
	
	We first consider sufficiently high energies $E$ of the quasiparticles, at which the self-energies of the
	quasiparticle  Green's function can be neglected in comparison with the energy $E$.
	Performing the summation of the ladder diagrams in Figs.~\ref{fig:quasiparticlerenormalisation}a and \ref{fig:quasiparticlerenormalisation}b
	(see \ref{sec:AppendixLadders} for details)
	gives the renormalised values of the disorder couplings (``impurity lines'') in the ``Cooper'' and ``excitonic'' channels:
	\begin{align}
		\label{gC}
		{g}_{c} \approx & g_0 \holOne \otimes \holOne 
		+ 2g_0\frac{ \left(\frac{p_0 g_0}{4\pi v^2}\ln \frac{vK}{\left| E \right|}\right)^2 }{1- \left(\frac{p_0g_0}{2\pi v^2} \ln \frac{vK}{\left| E \right|} \right)^2} \left(\holOne \otimes \holOne + \hat{\sigma}_{y}\otimes \hat{\sigma}_{y}\right)  \nonumber\\
		&
		+g_0\frac{\frac{p_0 g_0}{4\pi v^2} \ln \frac{vK}{\left| E \right|}}{1-\left(\frac{p_0g_0}{2\pi v^2} \ln \frac{vK}{\left| E \right|} \right)^2}
		\left(\hat{\sigma}_{x} \otimes \hat{\sigma}_{x} - \hat{\sigma}_{z}\otimes \hat{\sigma}_{z}\right);
	\end{align}
	
	\begin{align}
		\label{gE}
		{g}_{e} \approx &  g_0 \holOne \otimes \holOne 
		+ 2g_0\frac{ \left(\frac{p_0 g_0}{4\pi v^2}\ln \frac{vK}{\left| E \right|}\right)^2 }{1- \left(\frac{p_0g_0}{2\pi v^2} \ln \frac{vK}{\left| E \right|} \right)^2} \left(\holOne \otimes \holOne + \hat{\sigma}_{y}\otimes \hat{\sigma}_{y}\right) \nonumber\\
		&
		+g_0\frac{\frac{p_0 g_0}{4\pi v^2} \ln \frac{vK}{\left| E \right|}}{1-\left(\frac{p_0g_0}{2\pi v^2} \ln \frac{vK}{\left| E \right|} \right)^2}
		\left(\hat{\sigma}_{x} \otimes \hat{\sigma}_{x} + \hat{\sigma}_{z}\otimes \hat{\sigma}_{z}\right),
	\end{align}
	where $g_0$ is the bare disorder strength (values of the ``impurity line'') in Figs.~\ref{fig:DisorderChannels}b and \ref{fig:DisorderChannels}c;
	the matrices $\holOne$, $\hat\sigma_{x}$, $\hat\sigma_{y}$ and $\hat\sigma_z$ act in the pseudospin space of the nodal semimetal;
	$p_0$ is the radius of the nodal ring (cf. Fig.~\ref{fig:NDLS}) and
	$v$ is the quasiparticle velocity near the nodal line. Hereinafter, we assume, for simplicity,
	that the bare disorder strength $g_0$ is the same in both channels,
	which corresponds to short-range-correlated quenched disorder.
	
	UV renormalisations of disorder correlators can alternatively be studied 
	using a renormalisation group (RG) approach, by repeatedly integrating out shells of highest momenta and renormalising quasiparticle properties at lower momenta. Perturbative one-loop RG flow
	equations for random quenched perturbations in a nodal-line semimetal
	of various symmetries have been derived in Ref.~\cite{WangNandkishore:disorderEffectNLS}.

	%%%%%%%%%%%%%%%%%%%%%%%%%%%%%%%%%%
	
	\subsection{Self-energy and scattering rate}
	
	The leading contributions of the discussed singular disorder coupling in the ``Cooper'' and ``excitonic'' channels
	to the quasiparticle self-energy are shown in Figs.~\ref{fig:quasiparticlerenormalisation}c and \ref{fig:quasiparticlerenormalisation}d. Due to the intersection of propagators in Figs.~\ref{fig:quasiparticlerenormalisation}c,
	its value is suppressed relative to the value of diagram~\ref{fig:quasiparticlerenormalisation}d. The summation of these diagrams gives the (retarded) self-energy
	(see \ref{sec:Renormalisation of the quasiparticle energy} for details)
	\begin{align}
		\label{SelfEnergyR}
		\Sigma^R \left(E,g_0\right)= 
		-\frac{p_0 g_0}{2\pi v^2}
		\frac{\ln \frac{vK}{\left| E \right|}}
		{1-\frac{p_0 g_0}{2\pi v^2} \ln \frac{vK}{\left| E \right|}}E
		- i \frac{p_0 g_0}{4v^2}\frac{\left| E \right| }
		{1-\frac{p_0 g_0}{2\pi v^2} \ln \frac{vK}{\left| E \right|}}.
	\end{align}
	The imaginary part of the self-energy $\text{Im}\,\Sigma^R=(2\tau)^{-1}$, given by the last term of Eq.~\eqref{SelfEnergyR}, determines the elastic
	scattering rate $1/\tau$ as
	\begin{align}
		\frac{1}{\tau}
		= & 
		\frac{p_0 g_0}{2v^2}
		\frac
		{\left| E \right|}
		{{1-\frac{p_0 g_0}{2\pi v^2}\ln \frac{vK}{\left| E \right|}}}.
		\label{ScatteringRate}
	\end{align}
	
	%%%%%%%%%%%%%%%%%%%%%%%%%%%%%%%%%%%%%%%%%%%%%%%%%%%%%
	
	\subsection{Singularity in the disorder strength and mobility edge}

	In what follows, we consider sufficiently weak disorder, which
	correspond to a small dimensionless disorder strength
	\begin{align}
		\label{BareDisorderStrength}
		\frac{p_0 g_0} {2\pi v^2} \ll 1.
	\end{align}
	The parameter \eqref{BareDisorderStrength} is on the order of the inverse Ioffe-Regel parameter $\left(|E|\tau\right)^{-1}$ at the UV-cutoff energy. 
	The smallness of this parameter
	ensures the absence of localisation of the quasiparticles 
	near the UV energy cutoff.

	The expressions for the renormalised disorder strengths
	\eqref{gC} and \eqref{gE}
	and the self-energy~\eqref{SelfEnergyR} have a singularity at
	the energy
	\begin{align}
		\label{EcThreshold}
		E_c \approx
		vK\exp\left({-\frac{2\pi v^2}{p_0g_0}}\right).
	\end{align}
	We emphasise that this scale is smaller than the 
	UV energy cutoff $vK$ for the weak disorder strength under consideration [cf. Eq.~\eqref{BareDisorderStrength}]. In the opposite
	limit of strong disorder, $\frac{p_0 g_0}{2\pi v^2}\gtrsim 1$,
	the quasiparticle DoS is strongly broadened by 
	disorder, and the behaviours of the density of states, scattering rate and other observables are non-singular.

	Due to the rapid growth of the renormalised disorder strength 
	and the quasiparticle scattering rate near the energy $E_c$, this energy scale also plays the role of the 
	mobility edge.
	When the energy $E$ approaches $E_c$, 
	the corrections to the self-energy Eq.~\eqref{SelfEnergyR} to the Green's function
	become large and the renormalised disorder strength is described by Eqs.~\eqref{gC} and \eqref{gE}
	with the replacement
	\begin{align}
		E \rightarrow \tilde{E}=\max \left[\left|E\right|, \left|\text{Re}\,\Sigma^R(E,g_0)\right|\right]\sgn E.
	\end{align}
	In the case of a sufficiently large parameter $v K/ \left|E\right|$, however, the difference between $E$ and $\tilde{E}$
	may be neglected in the argument of the logarithm in Eqs.~\eqref{gC}-\eqref{ScatteringRate}.
	
	\subsection{Infrared cutoff}
	
	Quasiparticle properties are significantly affected by the UV renormalisations only for sufficiently large energies, $|E|\gg 1/\tau$ (cf. also Fig.~\ref{fig:nodalringscattering}). At low energies, $|E|\lesssim1/\tau$, the quasiparticle density of states is significantly broadened by disorder
	and may be assumed to have the same order of magnitude.\label{rep} The density of states and the transport properties of the quasiparticles in this
	regime require descriptions by means of ``low-energy'' methods, such as the kinetic equation or a non-linear sigma-model, that utilise the UV-renormalised
	quasiparticle parameters as input.
	
	Due to the rapid growth of the scattering rate $1/\tau$ near the energy $E_c$ given by Eq.~\eqref{EcThreshold}, the crossover
	between the regimes of UV and low-energy transport occurs in a small vicinity of that energy.
	According to Eqs.~\eqref{ScatteringRate} and \eqref{BareDisorderStrength},
	the crossover condition $|E|\sim 1/\tau$ is reached when
	\begin{align}
		1-\frac{p_0 g_0} {2\pi v^2}\ln \frac{vK}{|\tilde{E}|}\sim \frac{p_0 g_0} {2v^2}.
		\label{IRcutoff}
	\end{align}

	%%%%%%%%%%%%%%%%%%%%%%%%%%%%%%%%%%%%%%%%%%%%%%%%%%%%%%%%%%%%%%%%%%%%%%%%%%%%%%%%%%%%%%%%%%%%%%%%
	
	\section{Scaling of the density of states}
	
	\label{Sec:DoS}
	
	In this section, we investigate the behaviour of the disorder-averaged density of states (DoS).
	The DoS is an observable that reflects the discussed disorder-driven singularities and is readily observed
	in, e.g., ARPES experiments.
	The disorder-averaged DoS is given by
	\begin{align}
		\rho(E)=-\frac{1}{\pi}\text{Im}\int \frac{d\bp}{(2\pi)^2}\frac{dk}{2\pi}
		G_{E}^R\left(k,\bp\right),
		\label{DoSgenericExpression}
	\end{align}
	where $G_{E}^R\left(k,\bp\right)$ is the disorder-averaged retarded Green's function at energy $E$.

\begin{figure}[h]
		\centering
		\includegraphics[width=0.8\linewidth]{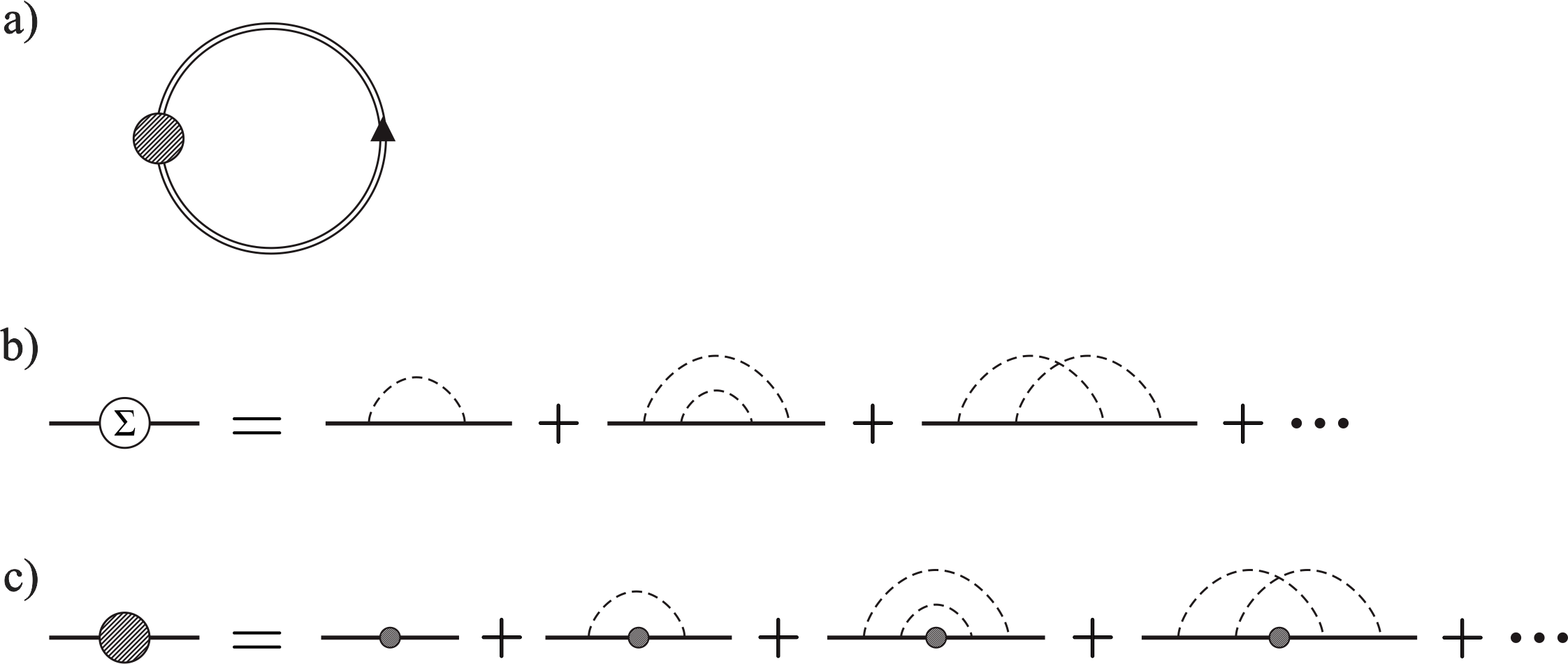}
		\caption{ 
			\label{fig:RenomalisedDOS}
			Diagrams for the renormalisation of the density of states.
			a) The renormalised density of states.
			b) The quasiparticle self-energy.
			c) The renormalisation of the density-of-states vertex by disorder (the bare vertex is given by $\rho_0=1$
			in momentum representation).
			Integration in diagrams b and c is carried out with respect to the UV momenta.
			The renormalisation of the DoS vertex is given by the same constant $\lambda$
			as the renormalisation of the energy term in the quasiparticle Green's function.
		}
	\end{figure}

	After performing the UV renormalisation of the disorder strength 
	and quasiparticle properties, the DoS can be reduced to~\cite{Syzranov:unconv} %(see also Appendix~\ref{sec:Density of states}) 
	\begin{align}
		\rho(E)=\lambda \rho_\text{clean} (\lambda E)
		\label{DoSgenericWeakDisorder}
	\end{align}
	for energies $|E|$ above the IR cutoff,
	where $\rho_\text{clean}=\frac{p_0|E|}{4\pi v^2}$ is the DoS in a disorder-free system,
	and 
	\begin{align}
		\label{lambda}
		\lambda=\left[E-\text{Re}\,\Sigma^R(E,g_0)\right]/E
	\end{align}
	is the UV ``energy renormalisation''. Equation~\eqref{DoSgenericWeakDisorder} can be understood as follows.
	The diagram for the DoS is shown in Fig.~\ref{fig:RenomalisedDOS}a and consists of the propagator 
	and the DoS vertex renormalised by the scattering through the UV momenta. The bare DoS vertex is given by $\rho=1$
	in the momentum representation. The renormalisation of this vertex, described by the diagrams in Figs.~\ref{fig:RenomalisedDOS}c,
	matches the renormalisation of the energy term in the Green's function, corresponding to the real part of the self-energy shown in Figs.~\ref{fig:RenomalisedDOS}b,
	which leads to a prefactor of $\lambda$ in Eq.~\eqref{DoSgenericWeakDisorder}.

	Utilising Eqs.~\eqref{SelfEnergyR} and \eqref{DoSgenericWeakDisorder}, we obtain the DoS above the energy
	$E_c$ [excluding a small vicinity of that energy given by Eq.~\eqref{IRcutoff}]:
	\begin{align}
		\rho_\text{high}=
		\frac{1}{\left(1-\frac{p_0 g_0}{2\pi v^2} \ln \frac{vK}{\left| E \right|}\right)^2}
		\frac{p_0}{4\pi v^2}\left| E \right|.
		\label{DoSmain1}
	\end{align}
	Equation~\eqref{DoSmain1} describes the DoS at energies and disorder strengths satisfying the condition
	$1- {p_0 g_0} \ln\left({vK}/{|\tilde{E}|}\right)/{2\pi v^2} \gg {p_0 g_0}/2v^2 $,
	i.e. before the crossover condition~\eqref{IRcutoff} is reached. At lower energies,
	the DoS has a constant order of magnitude due to the broadening by disorder.
	The value of such a low-energy DoS can be estimated using the conditions $\frac{p_0 g_0}{2\pi v^2} \ln (vK/|\tilde{E}|)\approx 1$  and Eq.~\eqref{IRcutoff}, which gives
	\begin{align}
		\rho_\text{low}\sim
		\frac{v^2}{\pi p_0 g_0^2}E_c
		\approx 
		\frac{v^3}{\pi p_0g^2_0}K
		%e^{-\frac{2\pi v^2}{p_0g_0}}.
		\exp\left({-\frac{2\pi v^2}{p_0g_0}}\right).
		\label{DoSmain2}
	\end{align}

	Equations~\eqref{DoSmain1} and \eqref{DoSmain2}
	are our main results for the DoS.
	The DoS displays a strong dependence on the disorder strength $g_0$ and displays a singular behaviour 
	summarised by Eq.~\eqref{DoSSummary}
	in a broad range of energies and disorder strengths, excluding a small region around the singularity:
	\begin{align}
		\rho(E,g_0)\propto \left[g_c(E)-g_0\right]^{-2}|E|,
		\label{SingularBehaviourRho}
	\end{align}
	where the critical disorder strength is given by $g_c(E)\approx 2\pi v^2/\left[p_0 \ln\left(\frac{vK}{|E|}\right)\right]$.
	Very close to the critical point, the singular behaviour~\eqref{SingularBehaviourRho} of the DoS
	is replaced by a constant dependence of the DoS on energy given by Eq.~\eqref{DoSmain2} [with the replacement $g_0\rightarrow g_c(E)$].

	%%%%%%%%%%%%%%%%%%%%%%%%%%%%%%%%%%%%%%%%%%%%%%%%%%%%%%%%%%%%%%%%%%%%%%%%%%%%%%%%%%%%%%%%%%%%%%%%%%%%%%%%%
	
	\section{Duality between the BCS instability in a metal and disorder-driven singularity	in a nodal-line semimetal}
	
	\label{sec:duality}
	
\begin{figure}[H]
		\centering
		\includegraphics[width=0.75\linewidth]{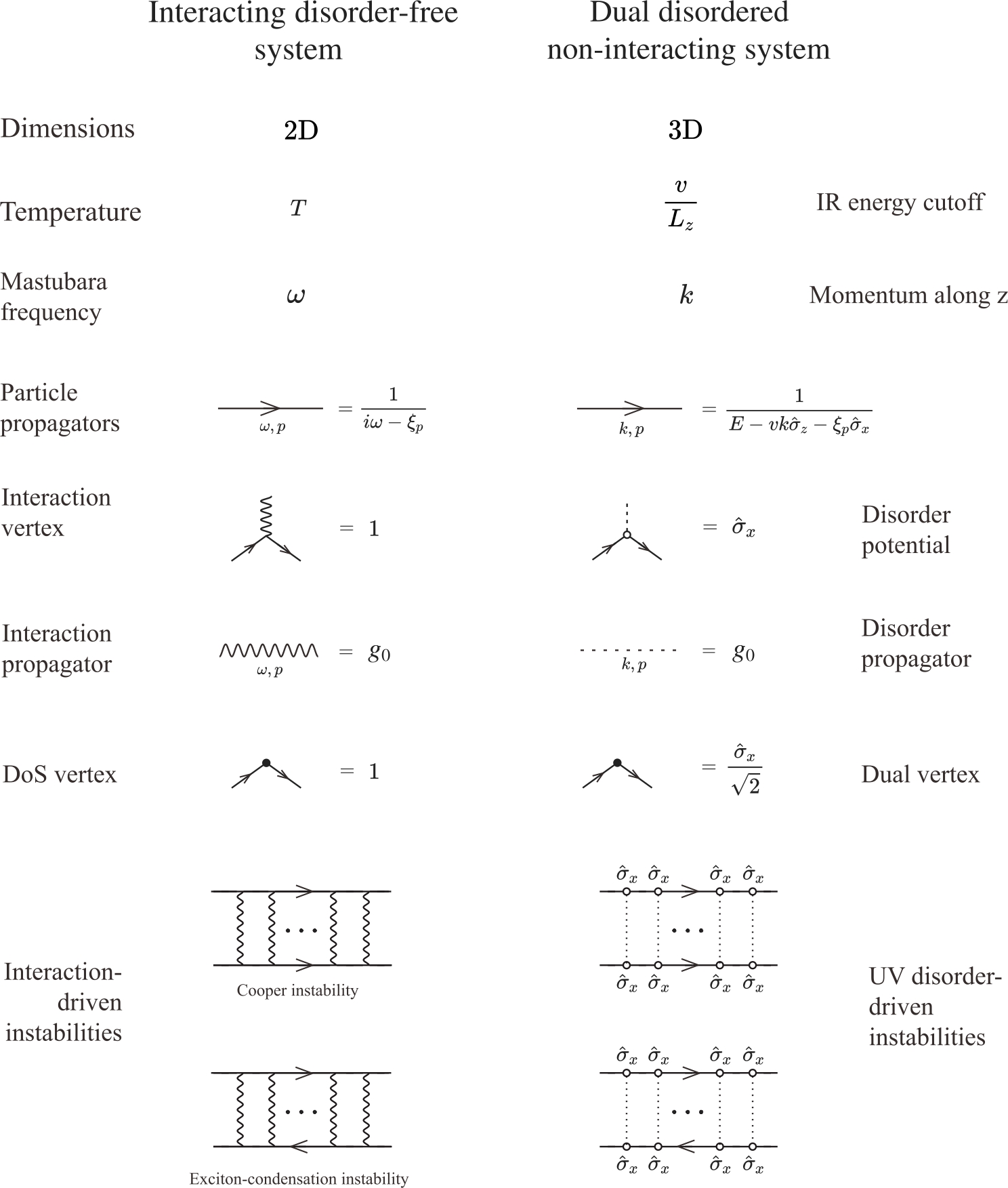}
		\caption{
			\label{fig:BCS-disorder}
			The correspondence between the diagrammatic elements of an interacting disorder-free system
			and an equivalent higher-dimensional disordered non-interacting system~\cite{SunSyzranov:duality}.
			The interacting system is described by means of the Matsubara diagrammatic technique.
			The disordered non-interacting system is described by the disorder-averaging diagrammatic technique~\cite{AGD}
			for particles with a (pseudo-)spin degree of freedom, which corresponds to the Pauli
			matrices $\sigma_{x}$ and $\sigma_z$.
			$\xi_\bp$ is the quasiparticle dispersion in the interacting system; $\omega$ is the Matsubara frequency; $k$
			is the momentum in the extra dimension of the disordered system.
		}
	\end{figure}

	It has recently been demonstrated in Ref.~\cite{SunSyzranov:duality} that 
	a $d$-dimensional interacting disorder-free system with a suppressed DoS at the chemical potential
	can be mapped to
	a $d+1$-dimensional system of non-interacting particles with quenched disorder.
	The mapping holds at all levels of the perturbation theory, with the respective dual elements of the diagrammatic technique shown in Fig.~\ref{fig:BCS-disorder}.
	In this mapping, the disordered system has an additional (pseudo-)spin structure corresponding to the Pauli matrices
	in Fig.~\ref{fig:BCS-disorder}.

	According to the duality transformation developed in Ref.~\cite{SunSyzranov:duality}, a 2D BCS model is equivalent to a 3D non-interacting disordered 
	system with the Hamiltonian given by 
	\begin{align}
		\hat{H} = v\left(|\bp|-p_0\right)\hat{\sigma}_{x}
		+vk\hat{\sigma}_{z} + u(\boldsymbol{\rho}) \hat{\sigma}_{x},
		\label{HdualBCS}
	\end{align}
	where $\xi_\bp=v\left(|\bp|-p_0\right)$ is the quasiparticle dispersion near the Fermi surface in the 2D BCS model;
	$p_0$ is the respective Fermi momentum and $u(\boldsymbol{\rho})$ is a short-range-correlated random potential.
	This model of a disordered nodal-line semimetal is rather similar to the nodal-line semimetal described by the 
	Hamiltonian~\eqref{H}. The only difference is the symmetry of disorder; while we consider scalar disorder
	in Eq.~\eqref{H}, the impurity-induced perturbations in the Hamiltonian~\eqref{HdualBCS} are proportional to $\hat\sigma_{x}$. As we have demonstrated in Sec.~\eqref{sec:UVrenormalisationMainResults}, UV renormalisation of 
	the potential disorder leads to the generation of other types of disorder, including that in the Hamiltonian~\eqref{HdualBCS}.

	This shows not only qualitative similarity between the disorder-driven instabilities studied here and the BCS/excitonic instabilities 
	in conventional metals but also establishes quantitative connection between them, with the exact duality achieved for certain disorder symmetries 
	($\hat V \propto \hat\sigma_{x}$).
	Performing the summation of the Cooper and exciton ladders for the model  
	described by the Hamiltonian~\eqref{HdualBCS}, we obtain
	renormalised disorder propagators (see ~\ref{sec:AppendixLadders_vector} for details):
	\begin{align}
		\tilde{g}_c=\frac{g_0}{2} \frac{\frac{p_0 g_0}{2\pi v^2}\ln\frac{vK}{|E|}}{1-\frac{p_0 g_0}{2\pi v^2}\ln\frac{vK}{|E|}}
		\left( \hat{\sigma}_{x} \otimes \hat{\sigma}_{x}- \hat{\sigma}_{z} \otimes \hat{\sigma}_{z} \right),
		\label{gCdual}
		\\
		\tilde{g}_e=\frac{g_0}{2}\frac{ \frac{p_0 g_0}{2\pi v^2}\ln\frac{vK}{|E|}}{1-\frac{p_0 g_0}{2\pi v^2}\ln\frac{vK}{|E|}}
		\left( \hat{\sigma}_{x} \otimes \hat{\sigma}_{x}+ \hat{\sigma}_{z} \otimes \hat{\sigma}_{z} \right).
		\label{gEdual}
	\end{align}
	The singularities of the renormalised couplings~\eqref{gCdual}
	and \eqref{gEdual} in the disordered nodal-line semimetal under consideration are
	dual to the singularities of the renormalised couplings near Cooper and exciton-condensation instabilities in interacting metals.
	The latter interaction-driven singularities manifest themselves, e.g., in the correlators of quasiparticle densities, which are dual to the correlators
	of the pseudospin density along the $x$ axis.

	The duality mapping in Ref.~\cite{SunSyzranov:duality} has been derived for a vanishing energy $E$
	of the particles in the disordered system. In this case, the low-energy (infrared)
	cutoff for the quasiparticles that contribute to the UV scattering processes is determined by the system size $L_z$
	along one of the directions (see Fig.~\ref{fig:BCS-disorder}). For the singularities described by 
	Eqs.~\eqref{gCdual} and \eqref{gEdual}, we replaced the respective cutoff by the quasiparticle energy $E$.

	For short system lengths 
	$L_z \ll v/|E|$, the energy $v/L_z$ serves as the infrared energy cutoff
	and replaces energy $E$ in Eqs.~\eqref{gCdual}-\eqref{gEdual},
	$|E|\rightarrow v/L_z$. In that regime, the DoS (and possibly other observables, e.g., conductivity)
	exhibit singularities as a function of $L_z$. Such a transition as a function of the length $L_z$
	is dual, in the sense of the duality mapping of Ref.~\cite{SunSyzranov:duality}, to
	the BCS temperature-driven transition.

	%%%%%%%%%%%%%%%%%%%%%%%%%%%%%%%%%%%%%%%%%%%%%%%%%%%%%%%%%%%%%%%%%%%%%%%%%%%%%%%%%%%%%%%%%%%%%%%%%%%%
	
	\section{Discussion and outlook}
	
	\label{Sec:conclusion}

\begin{figure}[h]
		\centering
		\includegraphics[width=0.3\linewidth]{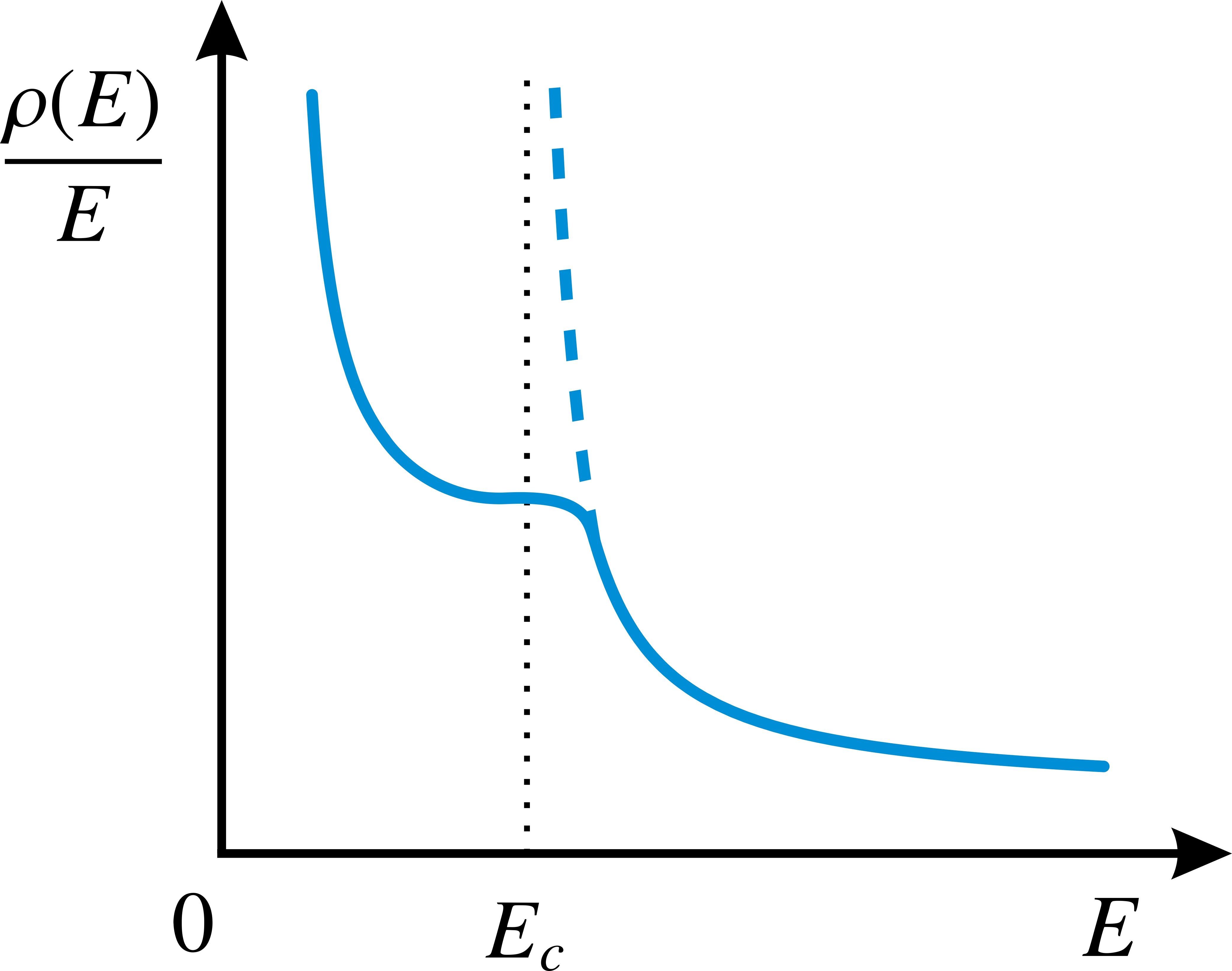}
		\caption{\label{fig:dos} The behaviour of the density of states as a function of energy.}
	\end{figure}

	We have studied the effects of quenched disorder on the density of states and quasiparticle properties in a nodal-line semimetal. 
	Significant renormalisations of the quasiparticle properties come from
	the ultraviolet scales, i.e. processes of scattering 
	through a large band of momenta whose width exceeds the 
	inverse mean free path $1/\ell$.
	This leads to a singular behaviour of multiple observables as a function of the disorder strength.
	We found that the density of states depends on the disorder strength $g_0$ and the quasiparticle energy
	$E$ as 
	\begin{align}
		\rho(E,g_0)\propto \left[g_c(E)-g_0\right]^{-2}|E|
		\label{DoSresultConclusion}
	\end{align}
	for $g_0$ exceeding the critical threshold $g_c(E)$, excluding a small vicinity of the threshold.
	In that vicinity, the singular behaviour~\eqref{DoSresultConclusion}
	crosses over to a constant as a function of $g_0$. 
	The behaviour of the quantity $\rho(E)/E$ as
	a function of energy $E$ is summarised in Fig.~\ref{fig:dos}. 
	Based on the scaling of the elastic scattering time $\tau$ and the density of states, we expect 
	the conductivity of a disordered nodal-line semimetal to show a singular behaviour of the form 
	\begin{align}
		\sigma \propto \tau \rho(E) \propto \left[g_c(E)-g_0\right]^{-1}/g_0,
		\label{ConductivityCritical}
	\end{align}
	although we leave a detailed microscopic investigation of the conductivity for future studies. 
	
	For certain disorder symmetries, a 3D disordered nodal-line semimetal considered in this paper can be mapped to 
	a model of a disorder-free 2D metal with attractive interactions using the duality mapping developed in Ref.~\cite{SunSyzranov:duality}. In such semimetals, disorder-driven instabilities are similar to 
	the Cooper and exciton-condensation instabilities. 
	The respective singular contributions to the disorder strength get generated, as a result of renormalisation
	of the properties of low-energy quasiparticles, for generic symmetries of disorder.
	Thus, although we have focused on two particular types of impurity symmetries in a nodal-line semimetal
	[cf. the Hamiltonians~\eqref{H} and \eqref{HdualBCS}], we still expect the predicted critical behaviour of observables
	to persist for arbitrary disorder.
	However, we will present a microscopic derivation of the density of states and conductivity for a generic disorder
	symmetry elsewhere.

	In this paper, we considered a model of a nodal-line semimetal in which all states on the nodal line have 
	the same energy. In realistic nodal-line semimetals, the energy of the nodal-line states may vary along the line.
	Our results apply if corrugations of the nodal line are significantly
	smaller than the energy scale $E_c$ at which the singularities 
	are observed.
	For stronger corrugations of the impurity line,
	we expect the critical scaling of observables as a function of the disorder strength $g_0$ to persist,
	although we leave a detailed microscopic investigation of such semimetals for future studies.

	{\bf Experimental observation.}
	In nodal-line semimetals, quenched disorder may be dominated by charged impurities.
	The potential of a charged impurity in a nodal-line semimetal has two characteristic lengths that significantly 
	exceed the inverse characteristic momentum $p_0^{-1}$ of the nodal line~\cite{SyzranovSkinner:NLSM}.
	Because the critical scaling of conductivity~\eqref{ConductivityCritical} requires short-range-correlated disorder,
	observing it in transport measurements requires the presence of uncharged defects, such
	as neutral impurities or vacancies.
	
	Observing the scaling of the density of states~\eqref{DoSresultConclusion}, however, is possible even if quenched 
	disorder comes from charged impurities because
	the singularity in the density of states $\rho(E)$ can be probed at energies $E$ away form the Fermi level $E_F$.
	Charged impurities are strongly screened and have a short screening radius 
	for sufficiently large energies $E_F$.
	The density of states $\rho(E)$ can then probed at $E\neq E_F$ in ARPES and STM experiments.

	%%%%%%%%%%%%%%%%%%%%%%%%%%%%%%%%%%%%%%%%%%%%%%%%%%%%%%%%%%%%%%%%%%%%%%%%%%%%%%%%%%%%%%%%%%%%%%%%%%%

	\clearpage
	
	\appendix

	\section{Cooper and excitonic channels for the renormalised disorder strength}
	
	\label{sec:AppendixLadders}
	
	In this section, we provide details of evaluating the diagrams in Figs.~\ref{fig:quasiparticlerenormalisation}. 
	As discussed in the main text, 
	the UV renormalisations come from the momenta far away from the Fermi surface, and the analytical properties of the Green's functions near the poles are not important for the respective contributions.
	In what follows, we consider, for simplicity, retarded Green's functions of the form %\ttr{I am not sure should we use Green's function instead of functions}
	$G(k,\bp)=\frac{\left(E+\frac{i}{2\tau}\right)+\xi_{\bf p}\hat{\sigma}_{x}+\xi_k\hat{\sigma}_{z}}{\left(E+\frac{i}{2\tau}\right)^2-\xi_{\bf p}^2-\xi_k^2 }$.
	In the case of short-range correlated, on which we focus in the main text, the bare couplings $g_0$ in the Cooper and excitonic channels match.

	\subsection{Cooper ladder}
	The one impurity-line-diagram in Figs.~\ref{fig:quasiparticlerenormalisation}a corresponds to $g_{c_1} = g_0 \left(\holOne \otimes \holOne \right)$. 
	Furthermore, the two-impurity-line diagram, which describes one step of the Cooper ladder, is given by
	\begin{align}
		\label{gc2.1}
		g_{c_2} =&   g_0^2\int \frac{d \bf p}{\left(2\pi\right)^2} \frac{dk}{2\pi}    G\left(k,\bf{p}\right)
		\otimes  G\left(-k,\bf{p}\right) 
		\nonumber\\ 
		=&   g_0^2\int \frac{d \bf p}{\left(2\pi\right)^2} \frac{dk}{2\pi}    
		\frac{\left(E+\frac{i}{2\tau}\right)+\xi_{\bf p}\hat{\sigma}_{x}+\xi_k\hat{\sigma}_{z}}{\left(E+\frac{i}{2\tau}\right)^2-\xi_{\bf p}^2-\xi_k^2 } 
		\otimes  \frac{\left(E+\frac{i}{2\tau}\right)+\xi_{\bf p}\hat{\sigma}_{x}-\xi_k\hat{\sigma}_{z}}{\left(E+\frac{i}{2\tau}\right)^2-\xi_{\bf p}^2-\xi_k^2 } 
		\nonumber\\ 
		\approx &   g_0^2\int \frac{d \bf p}{\left(2\pi\right)^2} \frac{dk}{2\pi}    \left\{
		\frac{\xi_{\bf p}^2\hat{\sigma}_{x} \otimes \hat{\sigma}_{x} - \xi_k^2 \hat{\sigma}_{z} \otimes \hat{\sigma}_{z}}{\left[\xi_{\bf p}^2+\xi_k^2 -\left(E+\frac{i}{2\tau}\right)^2\right]^2} 
		\right\}
		= \frac{g_0^2 p_0}{4\pi ^2 v^2} \int d \xi_p d \xi_k  \left\{
		\frac{\xi_{p}^2\hat{\sigma}_{x} \otimes \hat{\sigma}_{x} - \xi_k^2 \hat{\sigma}_{z} \otimes \hat{\sigma}_{z}}{\left[\xi_{p}^2+\xi_k^2 -\left(E+\frac{i}{2\tau}\right)^2\right]^2} 
		\right\},
	\end{align}
	where we have introduced $\xi_\bp=v\left(|\bp|-p_0\right)$ and $\xi_k=vk$, the contribution $\propto \left(E+\frac{i}{2\tau}\right)^2$ are neglected  due to the smallness of $E$ and $1/\tau$
	in comparison with $\xi_\bk$ and $\xi_p$, which are on the order of the UV energy scales.
	
	Introducing polar coordinates
	$r$ and $\theta$ via the relations $\xi_p = r cos\theta$ and $\xi_k = r sin\theta$ gives
	\begin{align}
		\label{gc2.2}
		g_{c_2}=&\frac{g_0^2 p_0}{4\pi ^2 v^2} \int_0 ^{vK}\frac{r^2 \cdot rdr}{\left[r^2 -\left(E+\frac{i}{2\tau}\right)^2\right]^2} \int_0 ^{2\pi} \cos^2 \theta d\theta \left(\hat{\sigma}_{x} \otimes \hat{\sigma}_{x} - \hat{\sigma}_{z} \otimes \hat{\sigma}_{z}\right)
		\nonumber\\ 
		\approx &\frac{g_0^2 p_0}{4\pi v^2} \left\{ \ln \frac{vK}{\max \left(\left|E\right|, 1/\tau\right)} + \frac{i}{2} \left[\frac{\pi}{2} + \arctan \left(\left|E\right| \tau - \frac{1}{4 \left|E\right| \tau}\right)\right] \sgn E \right\}
		\left(\hat{\sigma}_{x} \otimes \hat{\sigma}_{x} - \hat{\sigma}_{z} \otimes \hat{\sigma}_{z}\right).
	\end{align}
	In the limit $|E| \gg {1}/{\tau}$, 
	\begin{align}
		\label{gc2.4}
		g_{c_2}=& \frac{g_0^2 p_0}{4\pi v^2} \left(\ln \frac{vK}{\left| E \right|}+ \frac{\pi i \sgn E}{2} \right) 
		\left(\hat{\sigma}_{x} \otimes \hat{\sigma}_{x} - \hat{\sigma}_{z} \otimes \hat{\sigma}_{z}\right)
		\approx  \frac{1}{2}S g_0^2 \left(\hat{\sigma}_{x} \otimes \hat{\sigma}_{x} - \hat{\sigma}_{z} \otimes \hat{\sigma}_{z}\right),
	\end{align}
	where we introduce
	\begin{align}
		S = \frac{p_0}{2\pi v^2} \left(\ln \frac{vK}{\left| E \right|}+ \frac{\pi i \sgn E}{2} \right).
		\label{VariableS}
	\end{align}
	Keeping only the real part of $g_{c_2}$, which matters for the UV renormalisations, gives
	\begin{align}
		\label{VariableSprime}
		g_{c_2} 
		\approx & \frac{1}{2}S^{'} g_0^2  \left(\hat{\sigma}_{x} \otimes \hat{\sigma}_{x} -\hat{\sigma}_{z} \otimes \hat{\sigma}_{z} \right) ,
		\nonumber\\
		S^{'} = & \frac{p_0}{2\pi v^2} \ln \frac{vK}{\left| E \right|}.
	\end{align}
	The three-impurity-line diagram in Figs.~\ref{fig:quasiparticlerenormalisation}a can be expressed as
	\begin{align}
		\label{} 
		g_{c_3} =&   g_0^3\int \frac{d \bf p}{\left(2\pi\right)^2} \frac{d \bf p_1}{\left(2\pi\right)^2}  \frac{dk}{2\pi} \frac{dk_1}{2\pi}   G\left(k,\bf{p}\right) G\left(k_1,\bf{p_1}\right)
		\otimes  G\left(-k,\bf{p}\right) G\left(-k_1,\bf{p_1}\right)
		\nonumber\\ 
		=&   g_0^3\int \frac{d \bf p}{\left(2\pi\right)^2} \frac{d \bf p_1}{\left(2\pi\right)^2}  \frac{dk}{2\pi} \frac{dk_1}{2\pi}      
		\frac{\left(E+\frac{i}{2\tau}\right)+\xi_{\bf p}\hat{\sigma}_{x}+\xi_k\hat{\sigma}_{z}}{\left(E+\frac{i}{2\tau}\right)^2-\xi_{\bf p}^2-\xi_k^2 } 
		\frac{\left(E+\frac{i}{2\tau}\right)+\xi_{\bf p_1}\hat{\sigma}_{x}+\xi_{k_1}\hat{\sigma}_{z}}{\left(E+\frac{i}{2\tau}\right)^2-\xi_{\bf p_1}^2-\xi_{k_1}^2 } \nonumber\\
		& \otimes  \frac{\left(E+\frac{i}{2\tau}\right)+\xi_{\bf p}\hat{\sigma}_{x}-\xi_k\hat{\sigma}_{z}}{\left(E+\frac{i}{2\tau}\right)^2-\xi_{\bf p}^2-\xi_k^2 } 
		\frac{\left(E+\frac{i}{2\tau}\right)+\xi_{\bf p_1}\hat{\sigma}_{x}-\xi_{k_1}\hat{\sigma}_{z}}{\left(E+\frac{i}{2\tau}\right)^2-\xi_{\bf p_1}^2-\xi_{k_1}^2 } 
		\nonumber\\ 
		\approx &   g_0^3 \left(\frac{p_0}{4\pi v^2}\right)^2\int d \xi_{\bf p} d \xi_{\bf {p_1}} d \xi_k d \xi_{k_1}   
		\left\{
		\frac{\left(C^0_2 + C^2_2\right)\xi_{\bf p}^2\xi_{\bf p_1}^2  \holOne \otimes \holOne + C^1_2 \xi_{\bf p_1}^2 \xi_k^2 \hat{\sigma}_{y}\otimes \hat{\sigma}_{y}}{\left[\xi_{\bf p}^2+\xi_k^2 -\left(E+\frac{i}{2\tau}\right)^2\right]^2 \left[\xi_{\bf p_1}^2+\xi_{k_1}^2 -\left(E+\frac{i}{2\tau}\right)^2\right]^2} 
		\right\}
		\nonumber\\ 
		\approx & 2g_0 \left(\frac{1}{2}S^{'} g_0 \right)^2 \left(\holOne \otimes \holOne + \hat{\sigma}_{y}\otimes \hat{\sigma}_{y}\right) ,
	\end{align}
	where $C^k_n=\frac{n!}{k!(n-k)!}$ is the binomial coefficient.
	The four-impurity-line diagram in Figs.~\ref{fig:quasiparticlerenormalisation}a
	is given by
	\begin{align}
		\label{}
		g_{c_4} =&   g_0^4\int \frac{d \bf p}{\left(2\pi\right)^2} \frac{d \bf p_1}{\left(2\pi\right)^2}  \frac{d \bf p_2}{\left(2\pi\right)^2} \frac{dk}{2\pi} \frac{dk_1}{2\pi} \frac{dk_2}{2\pi}  G\left(k,\bf{p}\right) G\left(k_1,\bf{p_1}\right) G\left(k_2,\bf{p_2}\right)\otimes 
		G\left(-k,\bf{p}\right) G\left(-k_1,\bf{p_1}\right) G\left(-k_2,\bf{p_2}\right)
		\nonumber\\ 
		=&   g_0^4\int \frac{d \bf p}{\left(2\pi\right)^2} \frac{d \bf p_1}{\left(2\pi\right)^2}  \frac{d \bf p_2}{\left(2\pi\right)^2} \frac{dk}{2\pi} \frac{dk_1}{2\pi} \frac{dk_2}{2\pi}   \nonumber\\
		&\frac{\left(E+\frac{i}{2\tau}\right)+\xi_{\bf p}\hat{\sigma}_{x}+\xi_k\hat{\sigma}_{z}}{\left(E+\frac{i}{2\tau}\right)^2-\xi_{\bf p}^2-\xi_k^2 } 
		\frac{\left(E+\frac{i}{2\tau}\right)+\xi_{\bf p_1}\hat{\sigma}_{x}+\xi_{k_1}\hat{\sigma}_{z}}{\left(E+\frac{i}{2\tau}\right)^2-\xi_{\bf p_1}^2-\xi_{k_1}^2 } 
		\frac{\left(E+\frac{i}{2\tau}\right)+\xi_{\bf p_2}\hat{\sigma}_{x}+\xi_{k_2}\hat{\sigma}_{z}}{\left(E+\frac{i}{2\tau}\right)^2-\xi_{\bf p_2}^2-\xi_{k_2}^2 } \nonumber\\
		&\otimes  \frac{\left(E+\frac{i}{2\tau}\right)+\xi_{\bf p}\hat{\sigma}_{x}-\xi_k\hat{\sigma}_{z}}{\left(E+\frac{i}{2\tau}\right)^2-\xi_{\bf p}^2-\xi_k^2 } 
		\frac{\left(E+\frac{i}{2\tau}\right)+\xi_{\bf p_1}\hat{\sigma}_{x}-\xi_{k_1}\hat{\sigma}_{z}}{\left(E+\frac{i}{2\tau}\right)^2-\xi_{\bf p_1}^2-\xi_{k_1}^2 } 
		\frac{\left(E+\frac{i}{2\tau}\right)+\xi_{\bf p_2}\hat{\sigma}_{x}-\xi_{k_2}\hat{\sigma}_{z}}{\left(E+\frac{i}{2\tau}\right)^2-\xi_{\bf p_2}^2-\xi_{k_2}^2 } 
		\nonumber\\ 
		\approx &   g_0^4 \left(\frac{p_0}{4\pi v^2}\right)^3\int d \xi_{\bf p} d \xi_{\bf {p_1}} d \xi_{\bf {p_2}} d \xi_k d \xi_{k_1}   d \xi_{k_2}  
		\nonumber\\
		&\frac{\left(C^0_3 + C^2_3\right) \xi_{\bf p}^2\xi_{\bf p_1}^2\xi_{\bf p_2}^2 \hat{\sigma}_{x} \otimes \hat{\sigma}_{x} -
			\left(C^1_3 + C^3_3\right) \xi_{\bf p_1}^2 \xi_k^2\xi_{\bf p_2}^2  \hat{\sigma}_{z} \otimes \hat{\sigma}_{z}}{\left[\xi_{\bf p}^2+\xi_k^2 -\left(E+\frac{i}{2\tau}\right)^2\right]^2\left[\xi_{\bf p_1}^2+\xi_{k_1}^2 -\left(E+\frac{i}{2\tau}\right)^2\right]^2\left[\xi_{\bf p_2}^2+\xi_{k_2}^2 -\left(E+\frac{i}{2\tau}\right)^2\right]^2}
		\nonumber\\ 
		\approx & 4g_0 \left(\frac{1}{2}S^{'} g_0 \right)^3 \left(\hat{\sigma}_{x} \otimes \hat{\sigma}_{x} - \hat{\sigma}_{z}\otimes \hat{\sigma}_{z}\right).
	\end{align}
	Similarly, we obtain the values of any diagrams with $n$ ($n \geq 2$) impurity lines:
	\begin{align}
		\label{}
		g_{c_n} =
		\begin{cases}
			\frac{1}{2}g_0 \left(S^{'} g_0 \right)^{n-1} \left( \holOne \otimes \holOne+ \hat{\sigma}_{y} \otimes \hat{\sigma}_{y} \right)
			,&\text{when n is an odd number}
			\\
			\frac{1}{2}g_0 \left(S^{'} g_0 \right)^{n-1} \left( \hat{\sigma}_{x} \otimes \hat{\sigma}_{x}- \hat{\sigma}_{z} \otimes \hat{\sigma}_{z} \right), &\text{when n is an even number}.
		\end{cases}
	\end{align}
	Then the total value of the Cooper ladder in Figs.~\ref{fig:quasiparticlerenormalisation} is given by
	\begin{align}
		\label{}
		{g}_{c} \approx & g_0 \left(\holOne \otimes \holOne \right) + \frac{2g_0 \left(\frac{p_0g_0}{4\pi v^2}\ln \frac{vK}{\left| E \right|}\right)^2 }{1- \left(\frac{p_0g_0}{2\pi v^2} \ln \frac{vK}{\left| E \right|} \right)^2} \left(\holOne \otimes \holOne + \hat{\sigma}_{y}\otimes \hat{\sigma}_{y}\right)  %\nonumber\\
		+\frac{g_0 \frac{p_0g_0}{4\pi v^2} \ln \frac{vK}{\left| E \right|}}{1-\left(\frac{p_0g_0}{2\pi v^2} \ln \frac{vK}{\left| E \right|} \right)^2}
		\left(\hat{\sigma}_{x} \otimes \hat{\sigma}_{x} - \hat{\sigma}_{z}\otimes \hat{\sigma}_{z}\right).
	\end{align}
	
	%%%%%%%%%%%%%%%%%%%%%%%%%%%%%%%%%%%%%%%%%%%%%%%%%%%%%%%%%%5
	
	\subsection{``Excitonic'' ladder}
	
	The ``excitonic'' ladder in Figs.~\ref{fig:quasiparticlerenormalisation}b can be evaluated similarly to the 
	Cooper ladder in Figs.~\ref{fig:quasiparticlerenormalisation}a. 
	The first step of the ladder, i.e. the diagram with two impurity lines, is given by
	\begin{align}
		\label{}
		g_{e_2} 
		=&   g_0^2\int \frac{d \bf p}{\left(2\pi\right)^2} \frac{dk}{2\pi}    G\left(k,\bf{p}\right)
		\otimes  G\left(k,\bf{p}\right) 
		\nonumber\\ 
		=&   g_0^2\int \frac{d \bf p}{\left(2\pi\right)^2} \frac{dk}{2\pi}    
		\frac{\left(E+\frac{i}{2\tau}\right)+\xi_{\bf p}\hat{\sigma}_{x}+\xi_k\hat{\sigma}_{z}}{\left(E+\frac{i}{2\tau}\right)^2-\xi_{\bf p}^2-\xi_k^2 } 
		\otimes  \frac{\left(E+\frac{i}{2\tau}\right)+\xi_{\bf p}\hat{\sigma}_{x}+\xi_k\hat{\sigma}_{z}}{\left(E+\frac{i}{2\tau}\right)^2-\xi_{\bf p}^2-\xi_k^2 } 
		\nonumber\\ 
		\approx &   g_0^2\int \frac{d \bf p}{\left(2\pi\right)^2} \frac{dk}{2\pi}    \left\{
		\frac{\xi_{\bf p}^2\hat{\sigma}_{x} \otimes \hat{\sigma}_{x} + \xi_k^2 \hat{\sigma}_{z} \otimes \hat{\sigma}_{z}}{\left[\xi_{\bf p}^2+\xi_k^2 -\left(E+\frac{i}{2\tau}\right)^2\right]^2} 
		\right\}
		\approx  \frac{1}{2}S^{'} g_0^2 \left(\hat{\sigma}_{x} \otimes \hat{\sigma}_{x} +\hat{\sigma}_{z} \otimes \hat{\sigma}_{z} \right).
	\end{align}
	The value of the diagram with three impurity lines is given by
	\begin{align}
		\label{}
		g_{e_3} =&   g_0^3\int \frac{d \bf p}{\left(2\pi\right)^2} \frac{d \bf p_1}{\left(2\pi\right)^2}  \frac{dk}{2\pi} \frac{dk_1}{2\pi}   G\left(k,\bf{p}\right) G\left(k_1,\bf{p_1}\right)
		\otimes  G\left(k_1,\bf{p_1}\right) G\left(k,\bf{p}\right)
		\nonumber\\ 
		=&   g_0^3\int \frac{d \bf p}{\left(2\pi\right)^2} \frac{d \bf p_1}{\left(2\pi\right)^2}  \frac{dk}{2\pi} \frac{dk_1}{2\pi}      
		\frac{\left(E+\frac{i}{2\tau}\right)+\xi_{\bf p}\hat{\sigma}_{x}+\xi_k\hat{\sigma}_{z}}{\left(E+\frac{i}{2\tau}\right)^2-\xi_{\bf p}^2-\xi_k^2 } 
		\frac{\left(E+\frac{i}{2\tau}\right)+\xi_{\bf p_1}\hat{\sigma}_{x}+\xi_{k_1}\hat{\sigma}_{z}}{\left(E+\frac{i}{2\tau}\right)^2-\xi_{\bf p_1}^2-\xi_{k_1}^2 } \nonumber\\
		& \otimes  \frac{\left(E+\frac{i}{2\tau}\right)+\xi_{\bf p_1}\hat{\sigma}_{x}+\xi_{k_1}\hat{\sigma}_{z}}{\left(E+\frac{i}{2\tau}\right)^2-\xi_{\bf p_1}^2-\xi_{k_1}^2 } \frac{\left(E+\frac{i}{2\tau}\right)+\xi_{\bf p}\hat{\sigma}_{x}+\xi_k\hat{\sigma}_{z}}{\left(E+\frac{i}{2\tau}\right)^2-\xi_{\bf p}^2-\xi_k^2 } 
		\nonumber\\ 
		\approx &   g_0^3 \left(\frac{p_0}{4\pi v^2}\right)^2\int d \xi_{\bf p} d \xi_{\bf {p_1}} d \xi_k d \xi_{k_1}   
		\left\{
		\frac{\left(C^0_2 + C^2_2\right) \xi_{\bf p}^2\xi_{\bf p_1}^2 \holOne \otimes \holOne + C^1_2\xi_{\bf p_1}^2 \xi_k^2 \hat{\sigma}_{y}\otimes \hat{\sigma}_{y}}{\left[\xi_{\bf p}^2+\xi_k^2 -\left(E+\frac{i}{2\tau}\right)^2\right]^2 \left[\xi_{\bf p_1}^2+\xi_{k_1}^2 -\left(E+\frac{i}{2\tau}\right)^2\right]^2} 
		\right\}
		\nonumber\\ 
		\approx & 2g_0 \left(\frac{1}{2}S^{'} g_0 \right)^2 \left(\holOne \otimes \holOne + \hat{\sigma}_{y}\otimes \hat{\sigma}_{y}\right) ,
	\end{align}
	The contribution of the diagram with four impurity lines is given by
	\begin{align}
		\label{}
		g_{e_4} =&   g_0^4\int \frac{d \bf p}{\left(2\pi\right)^2} \frac{d \bf p_1}{\left(2\pi\right)^2}  \frac{d \bf p_2}{\left(2\pi\right)^2} \frac{dk}{2\pi} \frac{dk_1}{2\pi} \frac{dk_2}{2\pi}  G\left(k,\bf{p}\right) G\left(k_1,\bf{p_1}\right) G\left(k_2,\bf{p_2}\right)\otimes 
		G\left(k_2,\bf{p_2}\right) G\left(k_1,\bf{p_1}\right)  G\left(k,\bf{p}\right)
		\nonumber\\ 
		\approx&   g_0^4\int \frac{d \bf p}{\left(2\pi\right)^2} \frac{d \bf p_1}{\left(2\pi\right)^2}  \frac{d \bf p_2}{\left(2\pi\right)^2} \frac{dk}{2\pi} \frac{dk_1}{2\pi} \frac{dk_2}{2\pi}  \nonumber\\
		&\frac{\xi_{\bf p}\hat{\sigma}_{x}+\xi_k\hat{\sigma}_{z}}{\left(E+\frac{i}{2\tau}\right)^2-\xi_{\bf p}^2-\xi_k^2 } 
		\frac{\xi_{\bf p_1}\hat{\sigma}_{x}+\xi_{k_1}\hat{\sigma}_{z}}{\left(E+\frac{i}{2\tau}\right)^2-\xi_{\bf p_1}^2-\xi_{k_1}^2 } 
		\frac{\xi_{\bf p_2}\hat{\sigma}_{x}+\xi_{k_2}\hat{\sigma}_{z}}{\left(E+\frac{i}{2\tau}\right)^2-\xi_{\bf p_2}^2-\xi_{k_2}^2 } \nonumber\\
		&\otimes  
		\frac{\xi_{\bf p_2}\hat{\sigma}_{x}+\xi_{k_2}\hat{\sigma}_{z}}{\left(E+\frac{i}{2\tau}\right)^2-\xi_{\bf p_2}^2-\xi_{k_2}^2 } 
		\frac{\xi_{\bf p_1}\hat{\sigma}_{x}+\xi_{k_1}\hat{\sigma}_{z}}{\left(E+\frac{i}{2\tau}\right)^2-\xi_{\bf p_1}^2-\xi_{k_1}^2 } 
		\frac{\xi_{\bf p}\hat{\sigma}_{x}+\xi_k\hat{\sigma}_{z}}{\left(E+\frac{i}{2\tau}\right)^2-\xi_{\bf p}^2-\xi_k^2 } 
		\nonumber\\ 
		\approx &   g_0^4\int \frac{d \bf p}{\left(2\pi\right)^2} \frac{d \bf p_1}{\left(2\pi\right)^2}  \frac{d \bf p_2}{\left(2\pi\right)^2} \frac{dk}{2\pi} \frac{dk_1}{2\pi} \frac{dk_2}{2\pi}  \nonumber\\
		& \frac{\left(C^0_3 +C^2_3 \right) \xi_{\bf p}^2\xi_{\bf p_1}^2\xi_{\bf p_2}^2 \hat{\sigma}_{x} \otimes \hat{\sigma}_{x} +
			\left(C^1_3 +C^3_3 \right) \xi_{k}^2\xi_{k_1}^2\xi_{k_2}^2\hat{\sigma}_{z} \otimes \hat{\sigma}_{z}}{\left[\xi_{\bf p}^2+\xi_k^2 -\left(E+\frac{i}{2\tau}\right)^2\right]^2\left[\xi_{\bf p_1}^2+\xi_{k_1}^2 -\left(E+\frac{i}{2\tau}\right)^2\right]^2\left[\xi_{\bf p_2}^2+\xi_{k_2}^2 -\left(E+\frac{i}{2\tau}\right)^2\right]^2} 
		\nonumber\\ 
		\approx & 4g_0 \left(\frac{1}{2}S^{'} g_0 \right)^3 \left(\hat{\sigma}_{x} \otimes \hat{\sigma}_{x} + \hat{\sigma}_{z}\otimes \hat{\sigma}_{z}\right).
	\end{align}
	Similarly, we obtain the value of the diagram with $n$ (n $\geq 2$) impurity lines:
	\begin{align}
		\label{}
		g_{e_n} =
		\begin{cases}
			\frac{1}{2}g_0 \left(S^{'} g_0 \right)^{n-1} \left( \holOne \otimes \holOne+ \hat{\sigma}_{y} \otimes \hat{\sigma}_{y} \right)
			,&\text{when n is an odd number}
			\\
			\frac{1}{2}g_0 \left(S^{'} g_0 \right)^{n-1} \left( \hat{\sigma}_{x} \otimes \hat{\sigma}_{x} + \hat{\sigma}_{z} \otimes \hat{\sigma}_{z} \right), &\text{when n is an even number}.
		\end{cases}
	\end{align}
	Then value of the whole ``excitonic'' ladder in Figs.~\ref{fig:quasiparticlerenormalisation}b is given by
	\begin{align}
		\label{}
		{g}_{e} \approx &  g_0 \left(\holOne \otimes \holOne \right) + \frac{2g_0 \left(\frac{p_0g_0}{4\pi v^2}\ln \frac{vK}{\left| E \right|}\right)^2 }{1- \left(\frac{p_0g_0}{2\pi v^2} \ln \frac{vK}{\left| E \right|} \right)^2} \left(\holOne \otimes \holOne + \hat{\sigma}_{y}\otimes \hat{\sigma}_{y}\right) %\nonumber\\
		+\frac{g_0 \frac{p_0g_0}{4\pi v^2} \ln \frac{vK}{\left| E \right|}}{1-\left(\frac{p_0g_0}{2\pi v^2} \ln \frac{vK}{\left| E \right|} \right)^2}
		\left(\hat{\sigma}_{x} \otimes \hat{\sigma}_{x} + \hat{\sigma}_{z}\otimes \hat{\sigma}_{z}\right).
	\end{align}
	
	\section{Self-energy}
	\label{sec:Renormalisation of the quasiparticle energy}

	In this section, we provide details on the calculation of the UV contributions to the quasiparticle self-energy.
	The leading contributions of the Cooper channel to the self-energy, shown in Figs.~\ref{fig:quasiparticlerenormalisation}c,
	are suppressed compared to the ``excitonic'' contributions in Figs.~\ref{fig:quasiparticlerenormalisation}d by the 
	small parameter~\eqref{SmallParameterNodalLine}.
	
	The first-order contribution to the self-energy in Figs.~\ref{fig:quasiparticlerenormalisation}d is given by
	\begin{align}
		\label{} 
		E_{e_1} =& g_0\int \frac{d \bf p}{\left(2\pi\right)^2} \frac{dk}{2\pi}   G^R_0\left(k,\bf{p}\right) 
		\approx  -g_0 S E.
	\end{align}
	The next-order contribution is given by
	\begin{align}
		\label{} 
		E_{e_2} =& g^2_e\int \frac{d \bf p}{\left(2\pi\right)^2} \frac{dk}{2\pi} 
		\frac{d \bf p_1}{\left(2\pi\right)^2} \frac{dk_1}{2\pi}   G^R_0\left(k_1,\bf{p_1}\right) G^R_0\left(k,\bf{p}\right)  G^R_0\left(k_1,\bf{p_1}\right)
		\nonumber\\
		=& g^2_e\int \frac{d \bf p}{\left(2\pi\right)^2} \frac{dk}{2\pi} 
		G^R_0\left(k,\bf{p}\right)
		\int \frac{d \bf p_1}{\left(2\pi\right)^2} \frac{dk_1}{2\pi}   G^R_0\left(k_1,\bf{p_1}\right)   G^R_0\left(k_1,\bf{p_1}\right)
		\nonumber\\
		\approx & -SE \frac{1}{2}g_0 \left(S^{'} g_0 \right)  \left(1+1\right)
		\approx  -g_0 SE \left(S^{'} g_0 \right).
	\end{align}
	Similarly, we find $n$-th order contribution to the self-energy to be given by
	\begin{align}
		\label{}
		E_{e_n} =
		-g_0SE \left(S^{'} g_0 \right)^{n-1}.
	\end{align}
	The total quasiparticle self-energy hence can be shown as
	
	\begin{align}
		\label{Sigma1}
		\Sigma^R=  -\frac{g_0SE}{1-g_0 S^{'}}
		= -\frac{p_0 g_0}{2\pi v^2}
		\frac{\ln \frac{vK}{\left| E \right|}}
		{1-\frac{p_0 g_0}{2\pi v^2} \ln \frac{vK}{\left| E \right|}}E
		- i \frac{p_0 g_0}{4v^2}\frac{\left| E \right|}
		{1-\frac{p_0 g_0}{2\pi v^2} \ln \frac{vK}{\left| E \right|}}.
	\end{align}
	Utilising Eq.~\eqref{Sigma1}, we obtain the constant $\lambda$ characterising the renormalisation of the energy term in the 
	quasiparticle Green's function [cf. Eq.~\eqref{lambda}]:
	\begin{align}
		\label{}
		\lambda= \frac{1}{1-\frac{p_0 g_0}{2\pi v^2} \ln \frac{vK}{\left| E \right|}}.
	\end{align}
	
	\section{Ladder diagrams for impurity vertices $\propto\hat\sigma_{x}$}
	
	\label{sec:AppendixLadders_vector}
	
	In this section, we provide the details of evaluating the ladder diagrams
	in the model of a nodal-line semimetal described by the Hamiltonian~\eqref{HdualBCS}, dual to the 
	BCS model. The first step of the Cooper ladder, i.e. the diagram with two impurity lines, is given by
	\begin{align}
		\label{}
		\tilde{g}_{c_2} =&   g_0^2\int \frac{d \bf p}{\left(2\pi\right)^2} \frac{dk}{2\pi}    \hat{\sigma}_{x} G_0\left(k,\bf{p}\right) \hat{\sigma}_{x}
		\otimes  \hat{\sigma}_{x} G_0\left(-k,\bf{p}\right)  \hat{\sigma}_{x}
		=  \frac{g_0^2 p_0}{4\pi ^2 v^2} \int d \xi_p d \xi_k  \left[
		\frac{\xi_{p}^2\hat{\sigma}_{x} \otimes \hat{\sigma}_{x} - \xi_k^2 \hat{\sigma}_{z} \otimes \hat{\sigma}_{z}}{\left(\xi_{p}^2+\xi_k^2 -E^2\right)^2} 
		\right]
		\nonumber\\
		= & \frac{1}{2}g_0 \left(S^{'} g_0 \right) \left( \hat{\sigma}_{x} \otimes \hat{\sigma}_{x}- \hat{\sigma}_{z} \otimes \hat{\sigma}_{z} \right).
	\end{align}
	The value of the diagram with three impurity lines is given by
	\begin{align}
		\label{} 
		\tilde{g}_{c_3} =&   g_0^3\int \frac{d \bf p}{\left(2\pi\right)^2} \frac{d \bf p_1}{\left(2\pi\right)^2}  \frac{dk}{2\pi} \frac{dk_1}{2\pi}   \hat{\sigma}_{x} G_0\left(k,\bf{p}\right) \hat{\sigma}_{x} G_0\left(k_1,\bf{p_1}\right) \hat{\sigma}_{x}
		\otimes  \hat{\sigma}_{x} G_0\left(-k,\bf{p}\right) \hat{\sigma}_{x} G_0\left(-k_1,\bf{p_1}\right) \hat{\sigma}_{x}
		\nonumber\\ 
		\approx &   g_0^3 \left(\frac{p_0}{4\pi v^2}\right)^2\int d \xi_{\bf p} d \xi_{\bf {p_1}} d \xi_k d \xi_{k_1}   
		\left[
		\frac{\left(C^0_2 +C^2_2 \right)\xi_{\bf p}^2\xi_{\bf p_1}^2 \hat{\sigma}_{x} \otimes \hat{\sigma}_{x} - C^1_2\xi_{\bf p_1}^2 \xi_k^2  \hat{\sigma}_{z}\otimes \hat{\sigma}_{z}}{\left(\xi_{\bf p}^2+\xi_k^2 -E^2\right)^2 \left(\xi_{\bf p_1}^2+\xi_{k_1}^2 -E^2\right)^2} 
		\right]
		\nonumber\\ 
		\approx & \frac{1}{2}g_0 \left(S^{'} g_0 \right)^2 \left( \hat{\sigma}_{x} \otimes \hat{\sigma}_{x}- \hat{\sigma}_{z} \otimes \hat{\sigma}_{z} \right),
	\end{align}
	where the quantity $S^\prime$ is given by Eq.~\eqref{VariableSprime}.
	Similarly, we obtain the value of the diagram with $n\geq 2$ impurity lines:
	\begin{align}
		\label{} 
		\tilde{g}_{c_n} = \frac{1}{2}g_0 \left(S^{'} g_0 \right)^{n-1} \left( \hat{\sigma}_{x} \otimes \hat{\sigma}_{x}- \hat{\sigma}_{z} \otimes \hat{\sigma}_{z} \right).
	\end{align}
	After summing up the contribution from all ladder diagrams, we obtain the 
	renormalised disorder propagator in "Cooper" channel which is given by Eq.~\eqref{gCdual}.
	Similarly, the renormalised propagator in the ``excitonic'' channel can be obtained, which is given
	by Eq.~\eqref{gEdual}.

%%%%%%%%%%%%%%%%%%%%%%%%%%%%%%%%%%%%%%%%%%%%%%%%%%%%%%%%%%%%%%%%%%%%%%%%%

%%%%%%%%%%%%%%%%%%%%%%%%%%%%%%%%%%%%%%%%%%%%%%%%%%%%%%%%%%%%%%%%%%%%%%%%%%%%%%%%%%%%%%%%%%%%%%%%%%%%
%%%%%%%%%%%%%Bibliography%%%%%%%%%%%%%%%%%%%%%%%%%%%%%%%%%%%%%%%%%%%%%%%%%%%%%%%%%%%%%%%%%%%%%%%%%%%

%%%%%%%%%%%%%%%%%%%%%%%%%%%%%%%%%%%%%%%%%%%%%%%%%%%%%%%%%%%%%%%%%%%%%%%%%%%%%%%%%%%%%%%%%%%%%%%%%%%%
%%%%%%%%%%%%%%%%%%%%%%%%%%%%%%%%%%%%%%%%%%%%%%%%%%%%%%%%%%%%%%%%%%%%%%%%%%%%%%%%%%%%%%%%%%%%%%%%%%%%

\end{document}